\documentclass[%
 amsmath,amssymb,
 aps,
]{revtex4-2}

\usepackage{graphicx}
\usepackage{dcolumn}
\usepackage{bm}

\usepackage{txfonts}
\usepackage{comment}
\usepackage{empheq}
\usepackage{cprotect}

\graphicspath{{./figures/}}
\usepackage{color}
\definecolor{violet}{rgb}{0.56, 0.0, 1.0}

\begin{document}

\preprint{APS/123-QED}

\title{Nonequilibrium thermodynamics of populations of weakly-coupled low-temperature-differential Stirling engines with synchronous and asynchronous transitions}

\author{Songhao Yin}
\email{4894715006@edu.k.u-tokyo.ac.jp}
\author{Hiroshi Kori}
\author{Yuki Izumida}
\affiliation{
Graduate School of Frontier Sciences, The University of Tokyo, Kashiwa 277-8561, Japan
}

\date{\today}

\begin{abstract}
This study developed the theory of nonequilibrium thermodynamics for populations of low-temperature-differential (LTD) Stirling engines weakly-coupled in a general class of networks to clarify the effects of synchronous and asynchronous transitions on the power and thermal efficiency. We first show that synchronous (asynchronous) transitions increase (decrease) the power and thermal efficiency of weakly-coupled LTD Stirling engines based on quasilinear response relations between formally defined thermodynamic fluxes and forces. After that, we construct a conceptual model satisfying the quasilinear response relations to give a physical interpretation of the changes in power and thermal efficiency due to synchronous and asynchronous transitions, and justify the use of this conceptual model. We then show that the conceptual model, rather than the quasilinear response relations, preserves the thermodynamic irreversibility of the original model and thus gives more accurate results than those using the quasilinear response relations. Finally, we compare the dynamics between the original and the conceptual models for two-engine systems and show that the conceptual models roughly preserve the dynamical characteristics leading up to the synchronous transitions, while some detailed dynamical structures are lost.

\end{abstract}

\maketitle


\section{Introduction}
To realize a sustainable society, it is indispensable to utilize heat energy from low-temperature heat resources around us. One possible technology that realizes such an utilization is a low-temperature-differential (LTD) Stirling engine, which rotates autonomously with a slight temperature difference \cite{Senft2010,Senft2000,kongtragool2003review}. A minimal model with only two variables has been proposed to explain the mechanism of the rotational motion driven by the temperature difference, where the dynamical equations are described as a nonlinear pendulum, and the rotational state is described as a limit cycle \cite{izumida2018nonlinear}. With a slight temperature difference and load torque acting on the crank equipped with the engine, the linear relations between formally defined thermodynamic fluxes and forces were formulated in the quasilinear response regime \cite{izumida2020quasilinear}, and the maximum efficiency was obtained using the theory for conventional linear irreversible heat engines \cite{van2005thermodynamic}.

It has been widely observed in natural and artificial systems that self-sustained oscillators synchronize their frequencies by interacting with each other \cite{kuramoto1984chemical,Pikovsky}. It was expected that this synchronization mechanism may be utilized to increase the power and thermal efficiency of interacting LTD Stirling engines. To see how synchronous transitions due to the interaction between engines affect power and thermal efficiency, an analysis based on generalized quasilinear response relations for two weakly-coupled LTD Stirling engines was presented in a recent paper by the present authors \cite{PhysRevResearch.5.043268}. It was demonstrated that both power and thermal efficiency change due to synchronous and asynchronous transitions by changing the coupling strength, and that the maximum power and thermal efficiency are achieved when the engines are synchronized. It was also found that the transitions in power and thermal efficiency are characterized by a hysteresis, which is caused by different types of bifurcations in the synchronous and asynchronous transitions. Other studies analyzing the motion of interacting Stirling engines can be found in \cite{siches2022inertialess, kada2014synchronization}.

The present study extends the analysis performed in \cite{PhysRevResearch.5.043268} to populations of LTD Stirling engines weakly-coupled in a general class of networks, and constructs the theory of nonequilibrium thermodynamics with synchronous and asynchronous transitions. In Section II, we give a mathematical model of weakly-coupled low-temperature differential Stirling engines. In Section III, we develop the theory of nonequilibrium thermodynamics for the coupled engines as the main results. We first perform the analysis using formally defined thermodynamic fluxes and forces and the quasilinear response relations established between them to show that synchronous (asynchronous) transitions increase (decrease) the power and thermal efficiency. After that, we construct a conceptual model that satisfies the quasilinear response relations to give a physical interpretation of the transitions in power and thermal efficiency, and justify the use of this conceptual model. We then show that the conceptual model, rather than the quasilinear response relations, preserves the thermodynamic irreverbility of the original model and thus gives more accurate results than those using the quasilinear response relations. In Section IV, we compare the dynamics between the original and the conceptual models for two-engine systems and show that the conceptual models roughly preserve the dynamical characteristics leading up to the synchronous transitions, while some detailed dynamical structures are lost. Finally, in Section V, we summarize the study and give the future outlook.

\section{Model}
\begin{figure*}[htbp]
\begin{center}
\includegraphics[width=150mm]{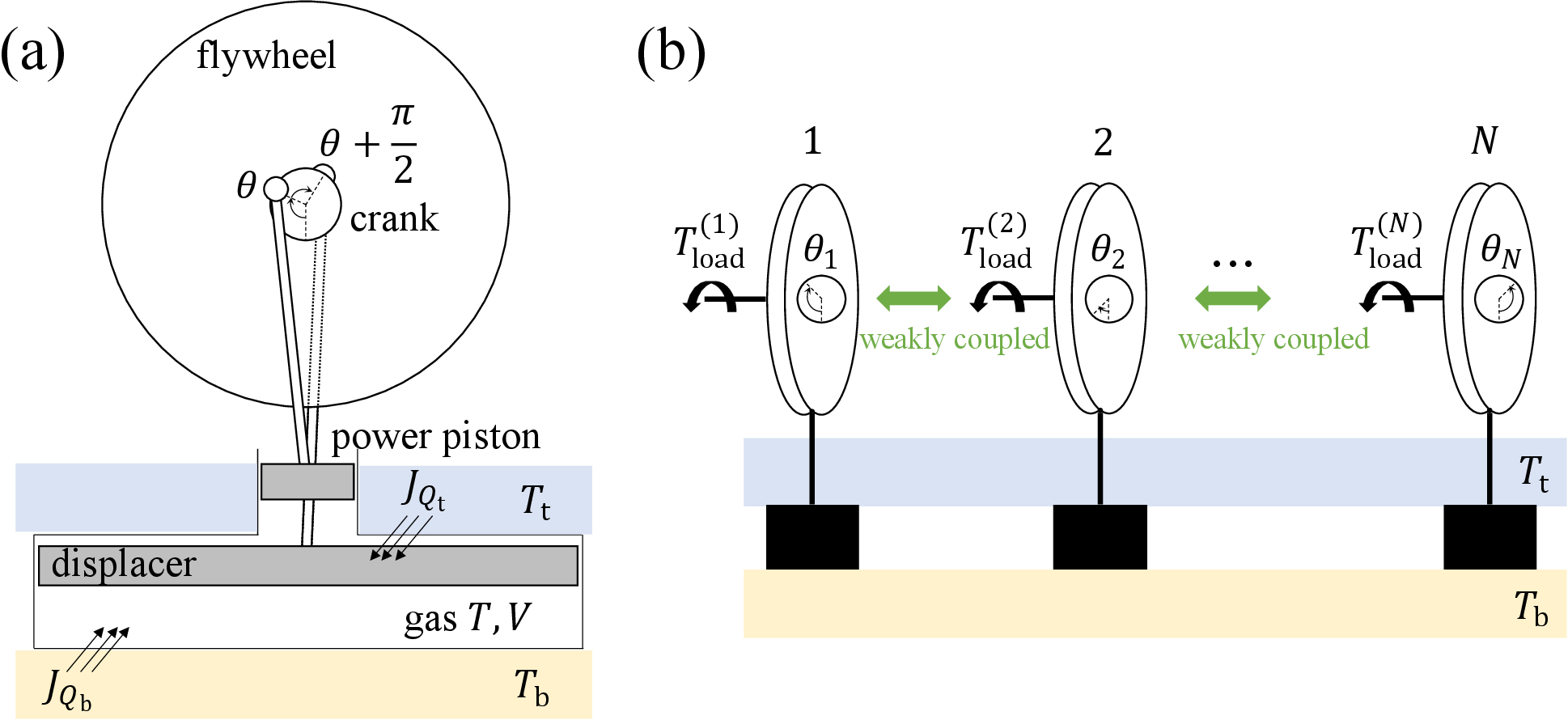}
\end{center}
\caption{(a) Front view of an LTD Stirling engine. The gas confined to the cylinders is in contact with the bottom and top heat reservoirs with temperatures $T_{\rm b}$ and $T_{\rm t}$. The displacer that advances the power piston in $\pi/2$ phase controls the coupling between the gas and the heat reservoirs. (b) Side view of $N$ LTD Stirling engines coupled in a chain with different load torques acting on the cranks.
}\label{fig.1}
\end{figure*}

A nondimensionalized minimal model of an LTD Stirling engine with only two variables has been proposed to study the rotational mechanism driven by the temperature difference $\Delta T$ of two heat reservoirs with temperature $T_{\rm b}=1+\Delta T/2$ and $T_{\rm t}=1-\Delta T/2$ ($\Delta T>0$) that are attached to the bottom and top surfaces of the large cylinders of the engine \cite{izumida2018nonlinear,izumida2020quasilinear} (Fig. \ref{fig.1} (a)): 
\begin{subequations}
\begin{equation}\label{single_dynamics0}
\frac{d\theta}{dt}=\omega,
\end{equation}
\begin{equation}\label{single_dynamics}
\frac{d\omega}{dt}=\sigma\left(\frac{T(\theta,\omega)}{V(\theta)}-P_{\rm air}\right)\sin\theta-\Gamma\omega-T_{\rm load}.
\end{equation}
\end{subequations}
Here, $\theta$ is the phase of the crank connected to the power piston, and $\omega$ is the angular velocity of that crank. A crank is also attached to the displacer, which is advanced by $\pi/2$ in phase compared to the crank attached to the power piston. These cranks are attached to a flywheel with a large moment of inertia to smoothen the rotational speed of the engine. The physical meanings of the parameters included in the above model are as follows: $\sigma$ is a positive constant determined by the surface areas of the large and small cylinders; $V(\theta)=2+\sigma(1-\cos\theta)$ and $\displaystyle T(\theta,\omega)=T_{\rm eff}(\theta)/\left(1+\frac{\sigma\sin\theta\omega}{GV(\theta)}\right)$ represent the volume and temperature of the gas confined to the cylinders, respectively; $\displaystyle T_{\rm eff}(\theta)=1+\frac{\sin\theta}{2}\Delta T$ is the effective temperature of the heat reservoirs that periodically changes depending on the phase; $G$ is the thermal conductance associated with the heat transfer between the gas and the surface of the large cylinder; $P_{\rm air}$ is the atmospheric pressure acting on the power piston; $\Gamma$ is the friction coefficient associated with the flywheel, and $T_{\rm load}>0$ is the load torque acting on the cranks. The minimal model was obtained by assuming a time-scale separation between the dynamics of the crank and the dynamics of the gas describing the energy conservation law, and the expression for the gas temperature $T(\theta,\omega)$ was obtained by the adiabatic elimination of the fast variable $T$, the gas temperature, in the differential equation describing the gas dynamics \cite{izumida2018nonlinear,izumida2020quasilinear}. The heat fluxes from the bottom and top surfaces of the large cylinder was assumed to obey the Fourier law
\begin{align}\label{heatflux_single}
J_{Q_{\rm m}}=G_{\rm m}(\theta)(T_{\rm m}-T(\theta,\omega)),
\end{align}
where $G_{\rm m}(\theta)$ with ${\rm m}={\rm b}$ (or t) represents the effective thermal conductance between the gas and the bottom (or top) heat reservoir. It was also assumed that $G_{\rm m}(\theta)\equiv G\chi_{\rm m}(\theta)$, where $\chi_{\rm m}(\theta)\hspace{1mm} (0\leq\chi_{\rm m}(\theta)\leq1)$ is a function that controls the coupling between the gas and the bottom or top heat reservoir, given as $\displaystyle\chi_{\rm b}(\theta)=\frac{1}{2}(1+\sin\theta)$ and $\displaystyle\chi_{\rm t}(\theta)=\frac{1}{2}(1-\sin\theta)$, reflecting the motion of the displacer. The dynamical equations (\ref{single_dynamics0}) and (\ref{single_dynamics}) describe the engine as a nonlinear pendulum, where the first term on the right-hand side of Eq. (\ref{single_dynamics}) represents the driven force due to the temperature difference. It has been experimentally verified that the minimal model (\ref{single_dynamics0}) and (\ref{single_dynamics}) correctly reflects the essential characteristics of a real LTD Stirling engine \cite{toyabe2020experimental}.

To investigate the synchronous and asynchronous transitions and the coupling effects on the power and thermal efficiency, we consider $N$ identical LTD Stirling engines that are weakly coupled, each with a different load torque applied, but placed between the same heat reservoirs, i.e., heat reservoirs at temperatures $T_{\rm b}$ and $T_{\rm t}$ (Fig. \ref{fig.1} (b)). The differential equations describing the motion of engine $i\hspace{1mm}(i=1, 2, \cdots, N)$ are given by
\begin{subequations}
\begin{equation}\label{dynamics0}
\frac{d\theta_i}{dt}=\omega_i,
\end{equation}
\begin{equation}\label{dynamics}
\frac{d\omega_i}{dt}=\sigma\left(\frac{T(\theta_i,\omega_i)}{V(\theta_i)}-P_{\rm air}\right)\sin\theta_i-\Gamma\omega_i-T_{\rm load}^{(i)}-\sum_{j\in\mathcal{N}_i}K_{ij}\sin(\theta_i-\theta_j),
\end{equation}
\end{subequations}
where the last term in Eq. (\ref{dynamics}) represents the coupling between engine $i$ and other engines with $\mathcal{N}_i$ being the set of engines that are coupled to engine $i$, and $K_{ij}=K_{ji}>0$ being the coupling strength between engine $i$ and engine $j$. We assume that the network structure representing the coupling of the engines is a connected one, and that all engines apply positive work to the load applied, i.e., $\omega_i>0$ for all $i\in\{1,2,\cdots,N\}$. As in Eq. (\ref{heatflux_single}), the heat flux $J_{Q_{\mathrm m}}^{(i)}$ (m=b, t) flowing into each engine $i$ is given by
\begin{align}\label{heatflux}
J_{Q_{\mathrm m}}^{(i)}=G_{\mathrm m}(\theta_i)(T_{\mathrm m}-T(\theta_i,\omega_i)).
\end{align}

\begin{figure*}[htbp]
\begin{center}
\includegraphics[width=100mm]{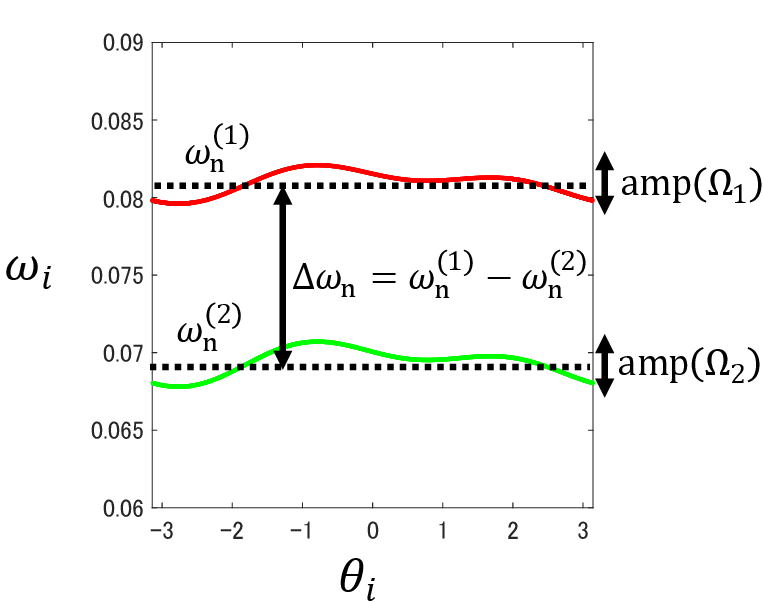}
\end{center}
\caption{Typical trajectories (red and green plots) of the limit cycle motion of the two engines satisfying condition (\ref{condition_nuturalfrequency_2engine}). The parameters of the engines were chosen as follows: $\sigma=0.02$, $\displaystyle p_{\mathrm{air}}=\frac{1}{V(\frac{\pi}{4})}=\frac{1}{2+\sigma(1-\cos\frac{\pi}{4})}\approx0.49854$, $G=1.5$, $\Gamma=0.001$, $\Delta T=1/29.3$, $T_{\mathrm{load}}^{(1)}=8.5324\times10^{-7}$, and $T_{\mathrm{load}}^{(2)}=1.2799\times10^{-5}$.
}\label{assump_illu}
\end{figure*}

Some important assumptions are made about the equations of motion (\ref{dynamics0}) and (\ref{dynamics}) describing the rotational motion and Eq. (\ref{heatflux}) describing the heat fluxes. We first assume that each engine is in the quasilinear response regime \cite{izumida2020quasilinear} in the absence of coupling, i.e., a limit cycle circling the surface of the phase space $\mathbb{T}\times\mathbb{R}$ that describes the rotational motion exists, and the effective frequency $\langle\omega_i\rangle$ and the averaged heat flux $\Bigl\langle J_{Q_{\rm b}}^{(i)}\Bigr\rangle$ from the heat reservoir of temperature 
$T_{\rm b}$ show a linear dependency w.r.t. $T_{\rm load}^{(i)}$ and $\Delta T$:
\begin{equation}\label{quasilinear_N_single}
\left[
  \begin{array}{c}
      \langle\omega_i\rangle \\     \Bigl\langle J_{Q_{\rm b}}^{(i)}\Bigr\rangle \\
  \end{array}
  \right]
  \approx
  \left[
  \begin{array}{cc}
      L_{11} & L_{12} \\
      L_{21} & L_{22}
    \end{array}
  \right]
  \left[
  \begin{array}{c}
      -T_{\rm load}^{(i)} \\
      \Delta T
  \end{array}
  \right].
\end{equation}
Here, $L_{11}, L_{12}, L_{21}, L_{22}$ are the quasilinear response coefficients and $\displaystyle\langle\cdots\rangle\equiv\lim_{\tau\to\infty}\frac{1}{\tau}\int_0^\tau\cdots dt$ denotes a long-time average, which is reduced to the average over one period for engines in periodic motion. In this regime, the periodic variation of the limit cycle of an uncoupled engine $i$ is sufficiently small compared to $|\omega_i|$, i.e.,
\begin{align}
{\rm amp}\left(\Omega_i\right)\ll|\omega_i|,
\end{align}
where $\displaystyle{\rm amp}\left(\Omega_i\right)\equiv\max_{s, s'\in\Omega_i}|s-s'|$ denotes the amplitude of the limit cycle and $\Omega_i$ denotes the set of $\omega_i$-components of the points on the limit cycle. The following conditions are also assumed to be satisfied for the natural frequency $\omega_{\mathrm n}^{(i)}$ of each engine:
\begin{align}\label{condition1_nuturalfrequency}
\Delta\omega_{\mathrm n}^{\mathrm{max}}\ll\omega_{\mathrm n}^{(i)},
\end{align}
\begin{align}\label{condition2_nuturalfrequency}
{\mathrm{amp}}\left(\Omega_i\right)\ll\Delta\omega_{\mathrm n}^{\mathrm{ave}}.
\end{align}
Here, the natural frequency $\omega_{\rm n}^{(i)}$ is given by $\omega_{\rm n}^{(i)}\equiv2\pi/T_i$, where $T_i$ denotes the period of the unperturbed limit cycle of engine $i$, $\Delta\omega_{\rm n}^{\rm max}\equiv\max_{i,j\in\{1,2,\cdots,N\}}|\omega_{\rm n}^{(i)}-\omega_{\rm n}^{(j)}|$ is the difference between the maximum and minimum natural frequencies, and $\displaystyle\Delta\omega_{\rm n}^{\rm ave}\equiv\frac{1}{N}\sum_{i=1}^N\left|\omega_{\rm n}^{(i)}-\frac{1}{N}\sum_{i=1}^N\omega_{\rm n}^{(i)}\right|$ is the average of the magnitude of the difference between the natural frequency and the natural frequency average. The conditions above ensure that the effects of higher-order terms in the derivation of the quasilinear response relations and averaged equations in the next section are negligible. A special case of this model with only two engines coupled was studied in Ref. \cite{PhysRevResearch.5.043268}, where the conditions regarding the natural frequencies reduce to
\begin{align}\label{condition_nuturalfrequency_2engine}
{\rm amp}\left(\Omega_i\right)\ll\Delta\omega_{\rm n}\ll\omega_{\rm n}^{(i)}.
\end{align}
Here, $\Delta\omega_{\rm n}\equiv|\omega_{\rm n}^{(1)}-\omega_{\rm n}^{(2)}|$ denotes the natural frequency difference of the two engines. Figure \ref{assump_illu} illustrates the typical trajectories of the limit cycle motion of the two engines that satisfy condition (\ref{condition_nuturalfrequency_2engine}).

\section{Main results: Theory of power and thermal efficiency based on nonequilibrium thermodynamics}
In this section, we develop a theory of nonequilibrium thermodynamics for weakly-coupled LTD Stirling engines with synchronous and asynchronous transitions. We first use the quasilinear response relations between thermodynamic fluxes and forces to show that the synchronous (asynchronous) transitions increase (decrease) the power and thermal efficiency in subsection A. After that, we construct a conceptual model satisfying the above quasilinear response relations to give a physical interpretation of the changes in power and thermal efficiency due to the weak coupling in subsection B. We then discuss the justification for using this conceptual model in subsection C. In subsection D, we perform numerical experiments to confirm the validity of the analysis on power and thermal efficiency. Finally, in subsection E, we review the analysis in subsections A $-$ C and show that the conceptual model, rather than the quasilinear response relations, preserves the thermodynamic irreversibility of the original model and thus gives more accurate results than those using the quasilinear response relations.

\subsection{Analysis based on the quasilinear response relations}

To obtain the thermodynamic fluxes and forces of the total thermodynamic system, we calculate the time-averaged entropy production rate of the total system $\displaystyle\biggl\langle\frac{d\sigma_{\mathrm{tot}}}{dt}\biggr\rangle$, which is the sum of the time-averaged entropy change rates of the two heat reservoirs since the time-averaged change in the entropy of the gas confined to the cylinder vanishes:
\begin{align}
\biggl\langle\frac{d\sigma_{\mathrm{tot}}}{dt}\biggr\rangle&=\sum_{i=1}^N
\left[ -\frac{\Bigl\langle J_{Q_{\rm b}}^{(i)}\Bigr\rangle}{T_{\rm b}}-\frac{\Bigl\langle J_{Q_{\rm t}}^{(i)}\Bigr\rangle-\Bigl\langle P_{\rm fric}^{(i)}\Bigr\rangle}{T_{\rm t}}
\right]\label{entropy_first}\\
&=\sum_{i=1}^N \left[-\frac{\Bigl\langle P_{\rm load}^{(i)}\Bigr\rangle+\Bigl\langle P_{K}^{(i)}\Bigr\rangle}{T_{\rm t}} 
+\Bigl\langle J_{Q_{\rm b}}^{(i)}\Bigr\rangle\left(\frac{1}{T_{\rm t}}-\frac{1}{T_{\rm b}}\right)\right]\label{entropy_conservation}\\
&\approx\sum_{i=1}^N\left[-\langle\omega_i\rangle T_{{\rm load}}^{(i)}-\sum_{j\in\mathcal{N}_i}K_{ij}\Bigl\langle\sin\left(\theta_i-\theta_j\right)\Bigr\rangle\langle\omega_i\rangle+\Bigl\langle J_{Q_{\rm b}}^{(i)}\Bigr\rangle\Delta T\right]\label{entropy_approx}\\
&=\langle\omega_{\rm m}\rangle\left(-\sum_{i=1}^{N}T_{\rm load}^{(i)}\right)+\sum_{i=1}^N\Bigl\langle\omega_{\rm d}^{(i)}\Bigr\rangle\left[\frac{1}{N}\left(\sum_{k=1}^{N}T_{\rm load}^{(k)}\right)-T_{\rm load}^{(i)}-\sum_{j\in\mathcal{N}_i}K_{ij}\Bigl\langle\sin\left(\theta_i-\theta_j\right)\Bigr\rangle\right]+\sum_{i=1}^N\Bigl\langle J_{Q_{\rm b}}^{(i)}\Bigr\rangle\Delta T\label{entropy_final}.
\end{align}
Here, $P_{\rm fric}^{(i)}\equiv\Gamma{\omega_i}^2$ is the power that is carried out against the friction torque, $P_{\rm load}^{(i)}\equiv T_{\rm load}^{(i)}\omega_i$ is the power that is carried out against the load torque, which is referred to as the brake power, $\displaystyle P_{K}^{(i)}\equiv \sum_{j\in\mathcal{N}_i}K_{ij}\sin\left(\theta_i-\theta_j\right)\omega_i$ is the power due to the weak coupling, $\displaystyle\langle\omega_{\rm m}\rangle\equiv\frac{1}{N}\sum_{i=1}^N\langle\omega_i\rangle$ is the mean effective frequency, $\Bigl\langle\omega_{\rm d}^{(i)}\Bigr\rangle\equiv\langle\omega_i\rangle-\langle\omega_{\rm m}\rangle$ is the difference of the effective frequency of engine $i$ from the mean effective frequency, and $\displaystyle\sum_{i=1}^N\Bigl\langle J_{Q_{\rm b}}^{(i)}\Bigr\rangle$ is the total heat flux from the heat reservoir of temperature $T_{\rm b}$. We have used the energy conservation law $\Bigl\langle J_{Q_{\rm b}}^{(i)}\Bigr\rangle+\Bigl\langle J_{Q_{\rm t}}^{(i)}\Bigr\rangle=\Bigl\langle P_{\rm load}^{(i)}\Bigr\rangle+\Bigl\langle P_{\rm fric}^{(i)}\Bigr\rangle+\Bigl\langle P_{K}^{(i)}\Bigr\rangle$ in Eq. (\ref{entropy_conservation}), approximated $\Bigl\langle P_{K}^{(i)}\Bigr\rangle$ as
\begin{align}\label{P_K_approx}
\Bigl\langle P_{K}^{(i)}\Bigr\rangle\approx\sum_{j\in\mathcal{N}_i}K_{ij}\Bigl\langle\sin\left(\theta_i-\theta_j\right)\Bigr\rangle\langle\omega_i\rangle
\end{align}
and approximated $T_{\rm b}$ and $T_{\rm t}$ as their mean value $1$ in Eq. (\ref{entropy_approx}), and used equations
\begin{align}\label{sum_omega_d}
\sum_{i=1}^N\Bigl\langle\omega_{\mathrm d}^{(i)}\Bigr\rangle=0,
\end{align}
\begin{align}\label{sum_coupling}
\sum_{i=1}^N\sum_{j\in\mathcal{N}_i}K_{ij}\Bigl\langle\sin\left(\theta_i-\theta_j\right)\Bigr\rangle=0
\end{align}
in Eq. (\ref{entropy_final}). Equation (\ref{P_K_approx}) is expected to be a good approximation given that conditions (\ref{quasilinear_N_single}) and (\ref{condition1_nuturalfrequency}) hold: the periodic fluctuations of $\omega_i$ are sufficiently small compared to $\langle\omega_i\rangle$ for uncoupled engines, and the synchronous transition occurs with sufficiently weak coupling strength so that the variation of $\omega_i$ is sufficiently small compared to $\langle\omega_i\rangle$ even on a long time scale for the coupled case.

Equation (\ref{entropy_final}) suggests that the terms $\displaystyle-\sum_{i=1}^{N}T_{\rm load}^{(i)}$, $\displaystyle\frac{1}{N}\left(\sum_{k=1}^{N}T_{\rm load}^{(k)}\right)-T_{\rm load}^{(i)}-\sum_{j\in\mathcal{N}_i}K_{ij}\Bigl\langle\sin\left(\theta_i-\theta_j\right)\Bigr\rangle$ and $\Delta T$ can be formally considered as thermodynamic forces with conjugate fluxes $\langle\omega_{\rm m}\rangle$, $\Bigl\langle\omega_{\rm d}^{(i)}\Bigr\rangle$, and $\displaystyle\sum_{i=1}^N\Bigl\langle J_{Q_{\rm b}}^{(i)}\Bigr\rangle$, respectively, for which the quasilinear response relations are obtained as follows:
\begin{equation}\label{quasilinear}
\left[
  \begin{array}{c}
      \bigl\langle\omega_{\rm m}\bigr\rangle \\     \displaystyle\sum_{i=1}^N\Bigl\langle J_{Q_{\rm b}}^{(i)}\Bigr\rangle \\
      \Bigl\langle\omega_{\rm d}^{(1)}\Bigr\rangle \\
      \Bigl\langle\omega_{\rm d}^{(2)}\Bigr\rangle \\
      \vdots \\
      \Bigl\langle\omega_{\rm d}^{(N)}\Bigr\rangle \\
  \end{array}
  \right]
  \approx
  \left[
  \begin{array}{cccccc}
      \displaystyle\frac{1}{N}L_{11} & L_{12} & 0 & 0 & \cdots & 0  \\
      L_{21} & NL_{22} & 0 & 0 & \cdots & 0  \\
      0 & 0 & L_{11} & 0 & \ddots & \vdots \\
      0 & 0 & 0 & L_{11} & \ddots & 0 \\
      \vdots & \vdots & \ddots & \ddots & \ddots & 0 \\
      0 & 0 & \cdots & 0 & 0 & L_{11}  \\
    \end{array}
  \right]
  \left[
  \begin{array}{c}
      -\displaystyle\sum_{i=1}^{N}T_{\rm load}^{(i)} \\
      \Delta T \\
      \displaystyle\frac{1}{N}\left(\sum_{k=1}^{N}T_{\rm load}^{(k)}\right)-T_{\rm load}^{(1)}-\sum_{j\in\mathcal{N}_1}K_{1j}\Bigl\langle\sin\left(\theta_1-\theta_j\right)\Bigr\rangle \\
      \displaystyle\frac{1}{N}\left(\sum_{k=1}^{N}T_{\rm load}^{(k)}\right)-T_{\rm load}^{(2)}-\sum_{j\in\mathcal{N}_2}K_{2j}\Bigl\langle\sin\left(\theta_2-\theta_j\right)\Bigr\rangle \\
      \vdots \\
      \displaystyle\frac{1}{N}\left(\sum_{k=1}^{N}T_{\rm load}^{(k)}\right)-T_{\rm load}^{(N)}-\sum_{j\in\mathcal{N}_N}K_{Nj}\Bigl\langle\sin\left(\theta_N-\theta_j\right)\Bigr\rangle \\
  \end{array}
  \right].
\end{equation}
Here, $L_{11}$, \hspace{1mm}$L_{12}$, \hspace{1mm}$L_{21}$, and \hspace{1mm}$L_{22}$ correspond to the quasilinear response coefficients in Eq. (\ref{quasilinear_N_single}) and are given by \cite{izumida2020quasilinear}
\begin{equation}
L_{11}=\frac{1}{\Gamma+\frac{\sigma^2}{G}\Bigl\langle\frac{\sin^2\theta}{V^2(\theta)}\Bigr\rangle_\theta},
\end{equation}
\begin{equation}
L_{12}=L_{21}=\frac{\frac{\sigma}{2}\Bigl\langle\frac{\sin^2\theta}{V(\theta)}\Bigr\rangle_\theta}{\Gamma+\frac{\sigma^2}{G}\Bigl\langle\frac{\sin^2\theta}{V^2(\theta)}\Bigr\rangle_\theta},
\end{equation}
\begin{equation}
L_{22}=\frac{G}{8}+\frac{\frac{\sigma^2}{4}\Bigl\langle\frac{\sin^2\theta}{V(\theta)}\Bigr\rangle_\theta^2}{\Gamma+\frac{\sigma^2}{G}\Bigl\langle\frac{\sin^2\theta}{V^2(\theta)}\Bigr\rangle_\theta},
\end{equation}
where $\displaystyle\langle\cdots\rangle_\theta\equiv\frac{1}{2\pi}\int_0^{2\pi}\cdots d\theta$ denotes the average w.r.t. the phase. The quasilinear response relations are obtained by taking the average of $J_{Q_{\rm b}}^{(i)}$ and both sides of Eq. (\ref{dynamics}), and neglecting higher-order terms in $\bigl\langle\omega_{\rm m}\bigr\rangle$, $\Bigl\langle\omega_{\rm d}^{(i)}\Bigr\rangle$, and $\Bigl\langle J_{Q_{\rm b}}^{(i)}\Bigr\rangle$ which come from the higher-order terms of the Taylor expansion of $T(\theta_i,\omega_i)$ and the terms resulting from approximating the effective frequency $\omega_i$ as a time-independent constant (See Appendix A for details).

The effects of synchronous and asynchronous transitions on the averaged brake power $\displaystyle\bigl\langle P_{\rm load}\bigr\rangle\equiv\sum_{i=1}^{N}\langle\omega_i\rangle T_{\rm load}^{(i)}$ and thermal efficiency $\displaystyle\eta\equiv\frac{\bigl\langle P_{\rm load}\bigr\rangle}{\sum_{i=1}^N\Bigl\langle J_{Q_{\rm b}}^{(i)}\Bigr\rangle}$ of the total system can be studied using the above quasilinear response relations. We first rewrite the averaged brake power $\bigl\langle P_{\rm load}\bigr\rangle$ in terms of fluxes $\bigl\langle\omega_{\rm m}\bigr\rangle$ and $\Bigl\langle\omega_{\rm d}^{(i)}\Bigr\rangle$:
\begin{align}
\bigl\langle P_{\rm load}\bigr\rangle
&=\bigl\langle\omega_{\rm m}\bigr\rangle\left(\sum_{i=1}^{N}T_{\rm load}^{(i)}\right)+\sum_{i=1}^{N}\Bigl\langle\omega_{\rm d}^{(i)}\Bigr\rangle T_{\rm load}^{(i)}\\
&=\bigl\langle P_{\rm m}\bigr\rangle+\bigl\langle P_{\rm rel}\bigr\rangle.
\end{align}
Here, $\displaystyle\bigl\langle P_{\rm m}\bigr\rangle\equiv\bigl\langle\omega_{\rm m}\bigr\rangle\left(\sum_{i=1}^{N}T_{\rm load}^{(i)}\right)$ denotes the power owing to the motion of the mean angle, and $\displaystyle\bigl\langle P_{\rm rel}\bigr\rangle\equiv\sum_{i=1}^{N}\Bigl\langle\omega_{\rm d}^{(i)}\Bigr\rangle T_{\rm load}^{(i)}$ denotes the power owing to the relative motion. To see the effects of synchronous and asynchronous transitions on the brake power, it is sufficient to investigate the change in $\bigl\langle P_{\rm rel}\bigr\rangle$ since $\bigl\langle\omega_{\rm m}\bigr\rangle$ is independent of the coupling from the quasilinear response relations. Substituting the quasilinear response relations w.r.t. fluxes $\Bigl\langle\omega_{\rm d}^{(i)}\Bigr\rangle$ in Eq. (\ref{quasilinear}), $\bigl\langle P_{\rm rel}\bigr\rangle$ can be rewritten as
\begin{align}
\bigl\langle P_{\rm rel}\bigr\rangle
&=\sum_{i=1}^{N}\Bigl\langle\omega_{\rm d}^{(i)}\Bigr\rangle\left[ T_{\rm load}^{(i)}-\frac{1}{N}\left(\sum_{k=1}^{N}T_{\rm load}^{(k)}\right)\right]\label{P_rel_0}. \\
&\approx L_{11}\sum_{i=1}^{N}\left[\frac{1}{N}\left(\sum_{k=1}^{N}T_{\rm load}^{(k)}\right)-T_{\rm load}^{(i)}-\sum_{j\in\mathcal{N}_i}K_{ij}\Bigl\langle\sin\left(\theta_i-\theta_j\right)\Bigr\rangle\right]\left[ T_{\rm load}^{(i)}-\frac{1}{N}\left(\sum_{k=1}^{N}T_{\rm load}^{(k)}\right)\right]\label{P_rel}.
\end{align}
Equation (\ref{P_rel}) implies that $\bigl\langle P_{\rm rel}\bigr\rangle$ takes a negative value that is proportional to the variance of load torque in the absence of coupling, and takes the value $0$ when the engines are synchronized by the weak coupling, which means that synchronous (asynchronous) transition increases (decreases) the averaged brake power, as in the case of two coupled engines \cite{PhysRevResearch.5.043268}. To see that the synchronized state achieves higher averaged brake power than any of the states leading up to the synchronous transition, we rewrite Eq. (\ref{P_rel_0}) as follows:
\begin{align}
\bigl\langle P_{\rm rel}\bigr\rangle
&=\sum_{i=1}^{N}\Bigl\langle\omega_{\rm d}^{(i)}\Bigr\rangle\left[ T_{\rm load}^{(i)}-\frac{1}{N}\left(\sum_{k=1}^{N}T_{\rm load}^{(k)}\right)\right]\label{P_rel1_0_0} \\
&\approx \sum_{i=1}^{N}\Bigl\langle\omega_{\rm d}^{(i)}\Bigr\rangle\left[ T_{\rm load}^{(i)}-\frac{1}{N}\left(\sum_{k=1}^{N}T_{\rm load}^{(k)}\right)+\sum_{j\in\mathcal{N}_i}K_{ij}\Bigl\langle\sin\left(\theta_i-\theta_j\right)\Bigr\rangle\right]\label{P_rel1_0} \\
&\approx -L_{11}^{-1}\sum_{i=1}^N\Bigl\langle\omega_{\rm d}^{(i)}\Bigr\rangle^2\label{P_rel1}.
\end{align}
We have added a term representing the averaged power due to the weak coupling of the $N$ coupled engines in Eq. (\ref{P_rel1_0}), which takes the value $0$ and is approximated as
\begin{align}\label{sum_P_coupling}
0=\Biggl\langle\sum_{i=1}^{N}\sum_{j\in\mathcal{N}_i}K_{ij}\sin\left(\theta_i-\theta_j\right)\omega_i\Biggr\rangle\approx\sum_{i=1}^{N}\sum_{j\in\mathcal{N}_i}K_{ij}\Bigl\langle\sin\left(\theta_i-\theta_j\right)\Bigr\rangle\langle\omega_i\rangle=\sum_{i=1}^{N}\sum_{j\in\mathcal{N}_i}K_{ij}\Bigl\langle\sin\left(\theta_i-\theta_j\right)\Bigr\rangle\Bigl\langle\omega_{\rm d}^{(i)}\Bigr\rangle,
\end{align}
and used the quasilinear response relations w.r.t. fluxes $\Bigl\langle\omega_{\rm d}^{(i)}\Bigr\rangle$ in Eq. (\ref{P_rel1}). The first equality in equation (\ref{sum_P_coupling}) can be easily derived by noting that $K_{ij}=K_{ji}$; the second approximation comes from Eq. (\ref{P_K_approx}) and the third equality can be obtained using Eq. (\ref{sum_coupling}). Since $\displaystyle\frac{1}{N}\sum_{i=1}^N\Bigl\langle\omega_{\rm d}^{(i)}\Bigr\rangle^2$ represents the dispersion of $\langle\omega_i\rangle$ around the mean value $\langle\omega_{\rm m}\rangle$, Eq. (\ref{P_rel1}) implies that the coupled system achieves higher averaged brake power with smaller rotational speed dispersion, and that the averaged brake power reaches the maximum when the engines are synchronized. On the other hand, the total heat flux from the heat reservoir of temperature $T_{\rm b}$ is independent of the coupling from the quasilinear response relations (\ref{quasilinear}). This implies that synchronous (asynchronous) transitions improve (deteriorate) both the averaged brake power and the thermal efficiency of the total system.

We have discussed the effects of synchronous and asynchronous transitions on power and thermal efficiency based on the quasilinear response relations (\ref{quasilinear}), which were derived by neglecting the higher-order terms representing the nonlinear dependence of the fluxes on the forces (c.f. Appendix A), but have not yet discussed the justification for neglecting the effects of these higher-order terms. Now we consider whether it is legitimate to use the quasilinear response relations to study the transitions in power and thermal efficiency. Let us first focus on the higher-order terms in fluxes $\langle\omega_{\rm m}\rangle$ and $\Bigl\langle\omega_{\rm d}^{(i)}\Bigr\rangle$. From conditions (\ref{quasilinear_N_single}) and (\ref{condition1_nuturalfrequency}), the deviation of $\omega_i$ during which $\theta_i$ changes by $2\pi$ is sufficiently small compared to the magnitude of $\langle\omega_i\rangle$, so the higher order terms can be neglected w.r.t. the effective frequency $\langle\omega_i\rangle$ for each engine, and therefore also w.r.t. the flux $\langle\omega_{\rm m}\rangle$. For the fluxes $\Bigl\langle\omega_{\rm d}^{(i)}\Bigr\rangle$, the higher order terms associated with $\Bigl\langle\omega_{\mathrm d}^{(i)}\Bigr\rangle$ may not be negligible compared to $\Bigl\langle\omega_{\mathrm d}^{(i)}\Bigr\rangle$, but if these higher-order terms are sufficiently small compared to the average of the variations of $\langle\omega_i\rangle$ due to the synchronous or asynchronous transitions, we can safely ignore them when focusing on the transitions in $\Bigl\langle\omega_{\rm d}^{(i)}\Bigr\rangle$. This is ensured by condition (\ref{condition2_nuturalfrequency}). Since the averaged brake power $\bigl\langle P_{\rm load}\bigr\rangle$ is the sum of the averaged power owing to the motion of the mean angle $\bigl\langle P_{\rm m}\bigr\rangle$ and the averaged power owing to the relative motion $\bigl\langle P_{\rm rel}\bigr\rangle$, we conclude that if the change in the averaged brake power is sufficiently larger than the sum of the higher-order terms in $\bigl\langle P_{\rm m}\bigr\rangle$ and $\bigl\langle P_{\rm rel}\bigr\rangle$, which comes from the higher-order terms in $\langle \omega_{\rm m}\rangle$ and $\Bigl\langle\omega_{\rm d}^{(i)}\Bigr\rangle$, then the quasilinear response relations (\ref{quasilinear}) w.r.t. $\langle \omega_{\rm m}\rangle$ and $\Bigl\langle\omega_{\rm d}^{(i)}\Bigr\rangle$ are well suited for studying the effects of synchronous and asynchronous transitions on the averaged brake power. This is ensured by conditions (\ref{quasilinear_N_single}), (\ref{condition1_nuturalfrequency}) and (\ref{condition2_nuturalfrequency}), which also justify the use of the quasilinear response relations (\ref{quasilinear}) when studying the effect of synchronous and asynchronous transitions on thermal efficiency.

Besides the quasilinear response relations, we also used an approximate formula (\ref{sum_P_coupling}) when discussing the optimality of the synchronous state, so the effect of the higher-order terms truncated in this approximation should also be discussed. As in the quasilinear response relations, the effect of these higher order terms can also be ignored given that conditions (\ref{quasilinear_N_single}), (\ref{condition1_nuturalfrequency}) and (\ref{condition2_nuturalfrequency}) are satisfied. In subsection E, we will discuss this approximation in more depth.

\subsection{Analysis based on the conceptual models constructed from the quasilinear response relations}

The reason that the fluxes $\bigl\langle\omega_{\rm m}\bigr\rangle$ and $\displaystyle\sum_{i=1}^N\Bigl\langle J_{Q_{\rm b}}^{(i)}\Bigr\rangle$ are independent of the coupling comes from the fact that the response coefficients
between the fluxes above and forces $\displaystyle\frac{1}{N}\left(\sum_{k=1}^{N}T_{\rm load}^{(k)}\right)-T_{\rm load}^{(i)}-\sum_{j\in\mathcal{N}_{i}}K_{ij}\Bigl\langle\sin\left(\theta_{i}-\theta_j\right)\Bigr\rangle$ are zero from Eq. (\ref{quasilinear}). The quasilinear response relations show a symmetric structure of response coefficients similar to Onsager symmetry in linear irreversible thermodynamics: the response coefficients between fluxes $\Bigl\langle\omega_{\rm d}^{(i)}\Bigr\rangle$ and forces $\displaystyle-\sum_{i=1}^{N}T_{\rm load}^{(i)}$, $\Delta T$ are also zero. Since both diagonal block matrices are symmetric and positive definite, and $\theta_i-\theta_j=\theta_{\rm d}^{(i)}-\theta_{\rm d}^{(j)}$ holds, where $\theta_{\rm d}^{(i)}\equiv\theta_i-\theta_{\rm m}$ and $\displaystyle\theta_{\rm m}\equiv\frac{1}{N}\sum_{i=1}^N\theta_i$, one of the systems satisfying the quasilinear response relations $(\ref{quasilinear})$ can be thought of as consisting of two isolated systems M and D, where the quasilinear response relations between thermodynamic fluxes and forces are given by
\begin{equation}\label{linear_M}
\left[
  \begin{array}{c}
      \bigl\langle\omega_{\rm m}\bigr\rangle \\     \Bigl\langle J_{Q_{\rm b}}^{\rm M}\Bigr\rangle \\
  \end{array}
  \right]
  \approx
  \left[
  \begin{array}{cc}
      \displaystyle\frac{1}{N}L_{11} & L_{12} \\
      L_{21} & NL_{22} \\
    \end{array}
  \right]
  \left[
  \begin{array}{c}
      -\displaystyle\sum_{i=1}^{N}T_{\rm load}^{(i)} \\
      \Delta T \\
  \end{array}
  \right],
\end{equation}
and
\begin{equation}\label{linear_D}
\left[
  \begin{array}{c}
      \Bigl\langle\omega_{\rm d}^{(1)}\Bigr\rangle \\
      \Bigl\langle\omega_{\rm d}^{(2)}\Bigr\rangle \\
      \vdots \\
      \Bigl\langle\omega_{\rm d}^{(N)}\Bigr\rangle \\
  \end{array}
  \right]
  \approx
  \left[
  \begin{array}{cccc}
      L_{11} & 0 & \cdots & 0 \\
      0 & L_{11} & \ddots & 0 \\
      \vdots & \ddots & \ddots & 0 \\
      0 & \cdots & 0 & L_{11}  \\
    \end{array}
  \right]
  \left[
  \begin{array}{c}
      \displaystyle\frac{1}{N}\left(\sum_{k=1}^{N}T_{\rm load}^{(k)}\right)-T_{\rm load}^{(1)}-\sum_{j\in\mathcal{N}_1}K_{1j}\Bigl\langle\sin\left(\theta_{\rm d}^{(1)}-\theta_{\rm d}^{(j)}\right)\Bigr\rangle \\
      \displaystyle\frac{1}{N}\left(\sum_{k=1}^{N}T_{\rm load}^{(k)}\right)-T_{\rm load}^{(2)}-\sum_{j\in\mathcal{N}_2}K_{2j}\Bigl\langle\sin\left(\theta_{\rm d}^{(2)}-\theta_{\rm d}^{(j)}\right)\Bigr\rangle \\
      \vdots \\
      \displaystyle\frac{1}{N}\left(\sum_{k=1}^{N}T_{\rm load}^{(k)}\right)-T_{\rm load}^{(N)}-\sum_{j\in\mathcal{N}_N}K_{Nj}\Bigl\langle\sin\left(\theta_{\rm d}^{(N)}-\theta_{\rm d}^{(j)}\right)\Bigr\rangle \\
  \end{array}
  \right],
\end{equation}
respectively. One of the systems satisfying Eq. $(\ref{linear_M})$ is the one where the differential equations w.r.t. $\theta_{\rm m}$ and the heat flux equation $J_{Q_{\rm b}}^{\rm M}$ are given by
\begin{empheq}[left={\empheqlbrace}]{alignat=2}
\hspace{1mm}&\frac{d\theta_{\rm m}}{dt}=\omega_{\rm m}\label{dynamics_M0}, \\
&\frac{d\omega_{\rm m}}{dt}+\left(\Gamma+\frac{\sigma^2}{G}\Biggl\langle\frac{\sin^2\theta}{V^2(\theta)}\Biggr\rangle_\theta\right)\omega_{\rm m}=-\frac{1}{N}\left(\sum_{i=1}^{N}T_{\rm load}^{(i)}\right)+\frac{\sigma}{2}\Biggl\langle\frac{\sin^2\theta}{V(\theta)}\Biggr\rangle_\theta\Delta T\label{dynamics_M},
\end{empheq}
and
\begin{align}
J_{Q_{\rm b}}^{\rm M}=N\left(\frac{G}{8}\Delta T+\frac{\sigma}{2}\Biggl\langle\frac{\sin^2\theta}{V(\theta)}\Biggr\rangle_\theta\omega_{\rm m}\right)\label{heatflux_M}
\end{align}
respectively, with the averaged entropy production rate $\biggl\langle\displaystyle\frac{d\sigma_{\rm M}}{dt}\biggr\rangle$ given by
\begin{align}
\biggl\langle\displaystyle\frac{d\sigma_{\rm M}}{dt}\biggr\rangle\approx\bigl\langle\omega_{\rm m}\bigr\rangle\left(-\sum_{i=1}^{N}T_{\rm load}^{(i)}\right)+\Bigl\langle J_{Q_{\rm b}}^{\rm M}\Bigr\rangle\Delta T\label{entropy_M}.
\end{align}
Equations (\ref{dynamics_M0})-(\ref{entropy_M}) indicate that system M can be considered as composed of the same heat reservoirs as those of the original system and a heat engine ${\rm M}_{\rm EG}$ working between the heat reservoirs with viscous damping coefficient of $\displaystyle N\left(\Gamma+\frac{\sigma^2}{G}\Biggl\langle\frac{\sin^2\theta}{V^2(\theta)}\Biggr\rangle_\theta\right)$ and load torque $\displaystyle\sum_{i=1}^{N}T_{\rm load}^{(i)}$. Indeed, in this case, by utilizing the energy conservation law
\begin{align}
\Bigl\langle J_{Q_{\mathrm b}}^{\mathrm M}\Bigr\rangle+\Bigl\langle J_{Q_{\mathrm t}}^{\mathrm M}\Bigr\rangle=\Bigl\langle P_{\mathrm{load}}^{\mathrm M}\Bigr\rangle+\Bigl\langle P_{\mathrm{fric}}^{\mathrm M}\Bigr\rangle,
\end{align}
the averaged entropy production rate $\biggl\langle\displaystyle\frac{d\sigma_{\mathrm M}}{dt}\biggr\rangle$ can be calculated as follows:
\begin{align}
\biggl\langle\displaystyle\frac{d\sigma_{\mathrm M}}{dt}\biggr\rangle
&=-\frac{\Bigl\langle J_{Q_{\mathrm b}}^{\mathrm M}\Bigr\rangle}{T_{\mathrm b}}-\frac{\Bigl\langle J_{Q_{\mathrm t}}^{\mathrm M}\Bigr\rangle-\Bigl\langle P_{\mathrm{fric}}^{\mathrm M}\Bigr\rangle}{T_{\mathrm t}}\\
&=-\frac{\Bigl\langle P_{\mathrm{load}}^{\mathrm M}\Bigr\rangle}{T_{\mathrm t}}+\Bigl\langle J_{Q_{\mathrm b}}^{\mathrm M}\Bigr\rangle\left(\frac{1}{T_{\mathrm t}}-\frac{1}{T_{\mathrm b}}\right)\\
&\approx\langle\omega_{\mathrm m}\rangle\left(-\sum_{i=1}^{N}T_{\mathrm{load}}^{(i)}\right)+\Bigl\langle J_{Q_{\mathrm b}}^{\mathrm M}\Bigr\rangle\Delta T\label{entropy_M_2}.
\end{align}
Here, $J_{Q_{\rm b}}^{\rm M}$ and $J_{Q_{\rm t}}^{\rm M}$ are the heat fluxes from the two heat reservoirs to the engine ${\rm M}_{\rm EG}$, $\displaystyle P_{\mathrm{load}}^{\mathrm M}\equiv\omega_{\mathrm m}\left(\sum_{i=1}^{N}T_{\mathrm{load}}^{(i)}\right)$ is the power that is carried out against the load torque, and $\displaystyle P_{\mathrm{fric}}^{\mathrm M}\equiv N\left(\Gamma+\frac{\sigma^2}{G}\biggl\langle\frac{\sin^2\theta}{V^2(\theta)}\biggr\rangle_\theta\right)(\omega_{\mathrm m})^2$ is the power that is carried out against the friction torque. The fact that Eq. (\ref{entropy_M_2}) agrees with Eq. (\ref{entropy_M}) indicates that the above interpretation is reasonable. Note that the specific physical picture of system M becomes clear only after incorporating the averaged entropy production rate (\ref{entropy_M}). Also note that the differential equation $(\ref{dynamics_M})$ with both sides multiplied by the same constant also satisfies the quasilinear response relation w.r.t. $\bigl\langle\omega_{\rm m}\bigr\rangle$, and the uncertainty due to this arbitrary constant is eliminated by the averaged entropy production rate (\ref{entropy_M}). The system satisfying Eq. $(\ref{linear_D})$, on the other hand, can be considered to be the one where the differential equations w.r.t. $\theta_{\rm d}^{(i)}$ are given by
\begin{empheq}
[left=\empheqlbrace]{align}
&\frac{d\theta_{\mathrm d}^{(i)}}{dt}=\omega_{\mathrm d}^{(i)}\label{dynamics_D0}, \\
&\frac{d\omega_{\mathrm d}^{(i)}}{dt}+\left(\Gamma+\frac{\sigma^2}{G}\Biggl\langle\frac{\sin^2\theta}{V^2(\theta)}\Biggr\rangle_\theta\right)\omega_{\mathrm d}^{(i)}=\frac{1}{N}\left(\sum_{k=1}^{N}T_{\mathrm{load}}^{(k)}\right)-T_{\mathrm{load}}^{(i)}-\sum_{j\in\mathcal{N}_i}K_{ij}\sin\left(\theta_{\mathrm d}^{(i)}-\theta_{\mathrm d}^{(j)}\right)\label{dynamics_D},
\end{empheq}
with the averaged entropy production rate $\biggl\langle\displaystyle\frac{d\sigma_{\rm D}}{dt}\biggr\rangle$ given by
\begin{align}
\biggl\langle\displaystyle\frac{d\sigma_{\rm D}}{dt}\biggr\rangle
&\approx\sum_{i=1}^N\Bigl\langle\omega_{\mathrm d}^{(i)}\Bigr\rangle\left[\frac{1}{N}\left(\sum_{k=1}^{N}T_{\mathrm{load}}^{(k)}\right)-T_{\mathrm{load}}^{(i)}-\sum_{j\in\mathcal{N}_i}K_{ij}\biggl\langle\sin\left(\theta_{\mathrm d}^{(i)}-\theta_{\mathrm d}^{(j)}\right)\biggr\rangle\right]\label{entropy_D}.
\end{align}
Equations (\ref{dynamics_D0})-(\ref{entropy_D}) indicate that system D can be considered as composed of a single heat reservoir of temperature $T_{\rm t}$, and $N$ coupled flywheels ${\rm D}_{\rm FW}^{(1)}, {\rm D}_{\rm FW}^{(2)},\cdots,{\rm D}_{\rm FW}^{(N)}$ that are in contact with that heat reservoir, each with viscous damping coefficient of $\displaystyle \Gamma+\frac{\sigma^2}{G}\Biggl\langle\frac{\sin^2\theta}{V^2(\theta)}\Biggr\rangle_\theta$ and load torque $T_{\rm load}^{(i)}-\displaystyle\frac{1}{N}\left(\sum_{k=1}^{N}T_{\rm load}^{(k)}\right)\hspace{1mm}(i=1, 2, \cdots, N)$, where the coupling network structure and the coupling strength are identical to those of the coupled Stirling engines. Indeed, in this case, the entropy production of system D is due to the dissipation of the heat generated by the viscous
friction caused by the rotation of these flywheels to the heat reservoir of temperature $T_{\rm t}$, with the entropy production rate $\displaystyle\frac{d\sigma_{\mathrm D}}{dt}$ given by
\begin{align}
\frac{d\sigma_{\mathrm D}}{dt}=\frac{P_{\mathrm{fric}}^{\mathrm D}}{T_{\mathrm t}}=\frac{1}{T_{\mathrm t}}\left(\Gamma+\frac{\sigma^2}{G}\Biggl\langle\frac{\sin^2\theta}{V^2(\theta)}\Biggr\rangle_\theta\right)\left(\sum_{i=1}^N\left(\omega_{\mathrm d}^{(i)}\right)^2\right)\approx\left(\Gamma+\frac{\sigma^2}{G}\Biggl\langle\frac{\sin^2\theta}{V^2(\theta)}\Biggr\rangle_\theta\right)\left(\sum_{i=1}^N\left(\omega_{\mathrm d}^{(i)}\right)^2\right),\label{entropy_D_t}
\end{align}
where $\displaystyle P_{\mathrm{fric}}^{\mathrm D}=\left(\Gamma+\frac{\sigma^2}{G}\Biggl\langle\frac{\sin^2\theta}{V^2(\theta)}\Biggr\rangle_\theta\right)\left(\sum_{i=1}^N\left(\omega_{\mathrm d}^{(i)}\right)^2\right)$ denotes the power carried out against the viscious friction, and $T_{\rm t}=1-\Delta T/2$ was approximated as 1 since $\Delta T\ll T_{\mathrm t}$. The averaged entropy production rate $\displaystyle\biggl\langle\frac{d\sigma_{\mathrm D}}{dt}\biggr\rangle$ can then be approximated as
\begin{align}
\biggl\langle\frac{d\sigma_{\mathrm D}}{dt}\biggr\rangle
&\approx\left(\Gamma+\frac{\sigma^2}{G}\Biggl\langle\frac{\sin^2\theta}{V^2(\theta)}\Biggr\rangle_\theta\right)\left(\sum_{i=1}^N\biggl\langle\left(\omega_{\mathrm d}^{(i)}\right)^2\biggr\rangle\right)\label{entropy_D_0}\\
&=\sum_{i=1}^N\Bigl\langle\omega_{\mathrm d}^{(i)}\Bigr\rangle\left[\frac{1}{N}\left(\sum_{k=1}^{N}T_{\mathrm{load}}^{(k)}\right)-T_{\mathrm{load}}^{(i)}\right]\label{entropy_D_1}\\
&\approx\sum_{i=1}^N\Bigl\langle\omega_{\mathrm d}^{(i)}\Bigr\rangle\left[\frac{1}{N}\left(\sum_{k=1}^{N}T_{\mathrm{load}}^{(k)}\right)-T_{\mathrm{load}}^{(i)}-\sum_{j\in\mathcal{N}_i}K_{ij}\biggl\langle\sin\left(\theta_{\mathrm d}^{(i)}-\theta_{\mathrm d}^{(j)}\right)\biggr\rangle\right]\label{entropy_D_2}.
\end{align}
Equation (\ref{entropy_D_1}) is valid since the time-averaged amount of heat generated by the viscous friction is equal to the time-averaged work done by the load torques on the flywheels. We have used the following approximation
\begin{align}\label{approx_0}
\sum_{i=1}^N\Bigl\langle\omega_{\mathrm d}^{(i)}\Bigr\rangle\sum_{j\in\mathcal{N}_i}K_{ij}\biggl\langle\sin\left(\theta_{\mathrm d}^{(i)}-\theta_{\mathrm d}^{(j)}\right)\biggr\rangle\approx0
\end{align}
in Eq. (\ref{entropy_D_2}). The validity of the above approximation comes from the fact that the loads $T_{\mathrm{load}}^{(i)}$ were chosen so that Eq. (\ref{P_rel1_0}) is a good approximation of Eq. (\ref{P_rel1_0_0}). The fact that Eq. (\ref{entropy_D_2}) agrees with Eq. (\ref{entropy_D}) indicates that the above interpretation is reasonable.

\begin{figure*}[htbp]
\begin{center}
\includegraphics[width=180mm]{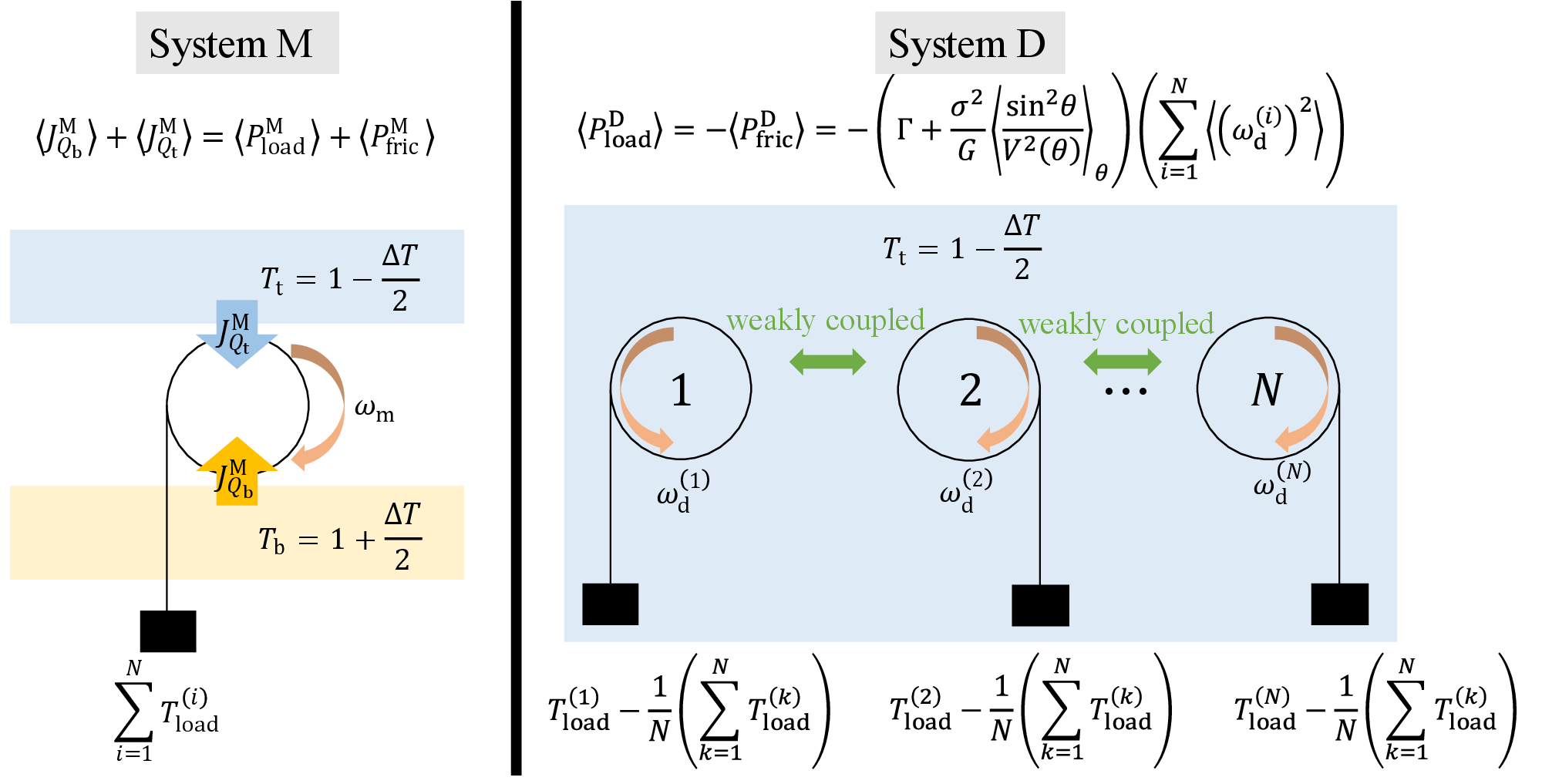}
\end{center}
\caption{Schematic diagram of two subsystems M and D constructed from the quasilinear response relations (\ref{quasilinear}) of $N$ LTD Stirling engines. System M consists of two heat reservoirs and a heat engine working between them. The heat engine performs positive work on the load acting on it. On the other hand, system D consists of a single heat reservoir and $N$ weakly coupled flywheels driven by different loads. All of the flywheels apply negative work to the loads acting on them. The yellow and blue areas represent the heat reservoirs of temperature $T_{\rm b}$ and $T_{\rm t}$, respectively. The signs of the loads in system D are represented by the positional relationship between the loads and the flywheels. The arrows in curved shape indicate the direction of rotation of the rotating bodies.
}\label{fig.physical_int}
\end{figure*}

Now we use the conceptual model constructed above to illustrate how weak coupling increases the averaged brake power and thermal efficiency of the total system (Note that the conceptual model is refered to as the physical picture of isolated systems M and D, not only the differential equations describing the rotational motion and the heat flux equation. This model was constructed through the quasilinear response relations (\ref{quasilinear}) and thus should reflects the thermodynamic irreversibility of the original coupled Stirling engine system). Figure \ref{fig.physical_int} shows a schematic diagram of systems M and D. The load acting on the engine ${\rm M}_{\rm EG}$ and flywheels ${\rm D}_{\rm FW}^{(i)}$ are represented by weights tied to the rotating bodies. Let us first consider the case where the original Stirling engines are uncoupled. In this case, flywheels ${\rm D}_{\rm FW}^{(i)}$ do not interact either and are simply driven by load torques of magnitude $\displaystyle\left|T_{\rm load}^{(i)}-\frac{1}{N}\left(\sum_{k=1}^{N}T_{\rm load}^{(k)}\right)\right|$, which correspond to the differences between the average load torque and the load torques on the original Stirling engines. Depending on the sign of the load torque, the flywheels in system D rotate in a positive or negative direction, but all flywheels perform negative work on the loads (i.e., the load acting on any flywheel applies positive work on the flywheel) regardless of the direction of rotation since the direction of the force exerted by the flywheel pulling the weight is opposite to the direction of the weight's motion, which is the essential difference from system M. In this sense, the engine ${\rm M}_{\rm EG}$ moves actively w.r.t the load, whereas the flywheels ${\rm D}_{\rm FW}^{(i)}$ are moved passively w.r.t the loads acting on them. When the motions of the flywheels settle to steady states, i.e., when the rotational speeds settle to constant values, the work done by the loads on the flywheels is completely converted to the heat due to viscous friction on the flywheels, which causes the entropy production in system D. If the original Stirling engines are coupled, flywheels ${\rm D}_{\rm FW}^{(i)}$ are also coupled with the same coupling strength and network structure. Unlike the uncoupled case, even if the motions of the flywheels settle to steady states, the rotational speeds are not constant values, so some of the work that the loads and the coupling forces perform to the flywheels are used to accelerate or decelerate the rotating bodies. However, the time-averaged work done by the loads on the flywheels $-\Bigl\langle P_{\mathrm{load}}^{\mathrm D}\Bigr\rangle$ is equal to the time-averaged work that is carried out against the friction torque $\Bigl\langle P_{\mathrm{fric}}^{\mathrm D}\Bigr\rangle$ (cf. Eq.(\ref{entropy_D_0}) and (\ref{entropy_D_1})). Since the coupling between the flywheels attempts to bring the variance of $\omega_{\rm d}^{(i)}$ close to zero, the negative power applied by the flywheels to the loads also approaches zero. Therefore, the averaged brake power of the total system is increased by the weak coupling, and at the same time, the total thermal efficiency is also improved since all the energy from the heat reservoir is absorbed by system M, which is independent of the coupling. When all flywheels in system D stop rotating, which corresponds to the fully synchronized state of the original Stirling engines, both the averaged brake power and the thermal efficiency of the total system reach the maximum values. We thus conclude that the increase of the averaged power and the thermal efficiency of the total system due to synchronous transitions is a result of the fact that the energy dissipation due to the viscous friction in system D is suppressed by synchronous transitions.

The conceptual model consisting of the two isolated systems M and D described above is not the only one that satisfies the quasilinear response relations $(\ref{quasilinear})$. One reason for the non-uniqueness of the model satisfying Eq. (\ref{quasilinear}) is that Eq. (\ref{quasilinear}) is given in the form of a long-time average, rather than on a time basis. For example, two subsystems that interact instantaneously but the interaction effect disappears on long time scales are also possible candidates for consideration. Since both the averaged power and thermal efficiency are evaluated by long-time averaging, the above non-uniqueness is not a substantial problem. Another reason for the non-uniqueness of the model satisfying Eq. (\ref{quasilinear}) is that the thermodynamic forces conjugate to the fluxes $\Bigl\langle\omega_{\mathrm d}^{(i)}\Bigr\rangle$ include the terms $\displaystyle-\sum_{j\in\mathcal{N}_i}K_{ij}\biggl\langle\sin\left(\theta_{\mathrm d}^{(i)}-\theta_{\mathrm d}^{(j)}\right)\biggr\rangle$, which depend on the dynamics of the original coupled engines. The conceptual model will incorrectly reflect not only the dynamical properties, but also the thermodynamic irreversibility of the original coupled system if the averaged coupling forces differ from those of the original coupled system since the thermodynamic forces, and thus the thermodynamic fluxes and the averaged entropy production rate are not preserved in this case. Therefore, the physical interpretation of the changes in power and thermal efficiency due to weak coupling for the constructed system cannot be directly applied to the coupled Stirling engines unless it is verified that the averaged coupling forces of the conceptual model constructed so far are approximately cocsistent numerically with those of the original coupled system. As we will see in the next subsection, the differential equations describing the rotational motion and the heat flux equation of the conceptual model can be obtained by averaging fast variables in those of the Stirling-engine model, and the higher-order terms produced by averaging fast variables can be neglected given that conditions (\ref{quasilinear_N_single}), (\ref{condition1_nuturalfrequency}) and (\ref{condition2_nuturalfrequency}) are satisfied, so the averaged coupling forces of the conceptual model are in close agreement with those of the original coupled system.

\subsection{Justification of the conceptual model}
In this section, we show that the differential equations describing the rotational motion of the conceptual model constructed in subsection B can be obtained by eliminating fast oscillating variables in those of the original coupled Stirling engine model. Since conditions (\ref{quasilinear_N_single}) and (\ref{condition1_nuturalfrequency}) ensure that the time scales of $\omega_i$ and $\theta_i-\theta_j$ are sufficiently large compared to that of $\theta_i$, i.e., $\omega_i$ and $\theta_i-\theta_j$ change only slightly during which $\theta_i$ varies by $2\pi$, if we replace the time derivative of $\omega_i$ by its time average during a one-round change in $\theta_i$, $\omega_i$ and $\theta_i-\theta_j$ can be regarded as constants so that we can just replace the driving force portion of the engine by its time average over one cycle of $\theta_i$ in Eq. (\ref{dynamics}):
\begin{align}
\frac{d\omega_i}{dt}
&\approx\Biggl\langle\frac{d\omega_i}{dt}\Biggr\rangle_{\rm rd} \\
&\approx\Biggl\langle\sigma\left(\frac{T(\theta_i,\omega_i)}{V(\theta_i)}-P_{\rm air}\right)\sin\theta_i\Biggr\rangle_{\rm rd}-\Gamma\omega_i-T_{\rm load}^{(i)}-\sum_{j\in\mathcal{N}_i}K_{ij}\sin\left(\theta_i-\theta_j\right) \\
&\approx\frac{\sigma}{2}\Biggl\langle\frac{\sin^2\theta}{V(\theta)}\Biggr\rangle_\theta\Delta T-\left(\Gamma+\frac{\sigma^2}{G}\Biggl\langle\frac{\sin^2\theta}{V^2(\theta)}\Biggr\rangle_\theta\right)\omega_i-T_{\rm load}^{(i)}-\sum_{j\in\mathcal{N}_i}K_{ij}\sin\left(\theta_i-\theta_j\right)\label{ave_omega}.
\end{align}
Here, $\langle\cdots\rangle_{\rm rd}$ denotes the time average during a one-round change in $\theta_i$. The time average of the driving force portion of each engine has been calculated when deriving the quasilinear response relations $(\ref{quasilinear})$, where $\omega_i$ was approximated as a constant, and can be calculated here in a similar way. Note that we have approximated $\langle\omega_i\rangle_{\rm rd}$ as $\omega_i$ in Eq. $(\ref{ave_omega})$. Ignoring the higher-order terms introduced by the averaging approximation, we obtain the following time-averaged equations:
\begin{subequations}
\begin{equation}\label{averaged0}
\frac{d\theta_i}{dt}=\omega_i,
\end{equation}
\begin{equation}\label{averaged}
\frac{d\omega_i}{dt}+\left(\Gamma+\frac{\sigma^2}{G}\Biggl\langle\frac{\sin^2\theta}{V^2(\theta)}\Biggr\rangle_\theta\right)\omega_i=\frac{\sigma}{2}\Biggl\langle\frac{\sin^2\theta}{V(\theta)}\Biggr\rangle_\theta\Delta T-T_{\rm load}^{(i)}-\sum_{j\in\mathcal{N}_i}K_{ij}\sin\left(\theta_i-\theta_j\right).
\end{equation}
\end{subequations}
It is then straightforward to obtain the dynamics w.r.t $(\theta_{\rm m}, \omega_{\rm m})$ and $(\theta_{\rm d}^{(i)}, \omega_{\rm d}^{(i)})$ from Eqs. (\ref{averaged0}) and (\ref{averaged}), which are given by Eqs. $(\ref{dynamics_M0})$ and $(\ref{dynamics_M})$ and Eqs. $(\ref{dynamics_D0})$ and $(\ref{dynamics_D})$. This means that the dynamics of the rotational motion in isolated subsystems M and D defined in the previous section can be obtained by averaging fast oscillating variables and ignoring higher-order terms. Conditions (\ref{quasilinear_N_single}) and (\ref{condition1_nuturalfrequency}) ensure that the effect due to truncated higher order terms are negligible when focusing on the dynamics w.r.t. $(\theta_{\mathrm m}, \omega_{\mathrm m})$, while condition (\ref{condition2_nuturalfrequency}) is also necessary to ensure that the effect of the higher order terms w.r.t. $(\theta_{\mathrm d}^{(i)}, \omega_{\mathrm d}^{(i)})$ are negligible. The conceptual model is thus expected to be reasonable as giving a physical interpretation of the changes in power and thermal efficiency due to weak coupling given that conditions (\ref{quasilinear_N_single}), (\ref{condition1_nuturalfrequency}) and (\ref{condition2_nuturalfrequency}) are satisfied. In Section IV, we will see that the analytically obtained values of the synchronous and asynchronous transition points of the averaged equations (\ref{averaged0})-(\ref{averaged}) are in good agreement with those of the original model (\ref{dynamics0})-(\ref{dynamics}) obtained by numerical experiments for the case of two coupled engines. The heat flux equation $(\ref{heatflux_M})$ in system M can be obtained similarly by replacing the heat flux $J_{Q_{\rm b}}^{(i)}(\theta_i,\omega_i)$ by its
time average during a one-round change in $\theta_i$ while considering the slow variable $\omega_i$ as a constant value:
\begin{align}
J_{Q_{\rm b}}^{(i)}\left(\theta_i,\omega_i\right)&\approx\Bigl\langle J_{Q_{\rm b}}^{(i)}(\theta_i,\omega_i)\Bigr\rangle_{\rm rd} \\
&=\Biggl\langle G\frac{1+\sin\theta_i}{2}\left(T_{\rm b}-T\left(\theta_i,\omega_i\right)\right)\Biggr\rangle_{\rm rd}\\
&\approx\frac{G}{8}\Delta T+\frac{\sigma}{2}\Biggl\langle\frac{\sin^2\theta}{V(\theta)}\Biggr\rangle_\theta\omega_i\label{J_b_ave_first}.
\end{align}
The higher-order terms produced by the averaging approximation of the heat flux can be ignored given that conditions (\ref{quasilinear_N_single}) and (\ref{condition1_nuturalfrequency}) are satisfied.

The above discussion confirms the validity of the conceptual model constructed in subsection B, but there are still several issues to be addressed. One is that the results of the analysis in subsection B using the conceptual model do not agree with the results of the analysis in subsection A using the quasilinear response relations (\ref{quasilinear}): the averaged power $\Bigl\langle P_{\mathrm{load}}^{\mathrm D}\Bigr\rangle$ applied by the flywheels ${\mathrm D}_{\mathrm FW}^{(1)}, {\mathrm D}_{\mathrm FW}^{(2)},\cdots,{\mathrm D}_{\mathrm FW}^{(N)}$ to the loads (or the averaged power $-\Bigl\langle P_{\mathrm{fric}}^{\mathrm D}\Bigr\rangle$ due to the viscous friction) is given by Eq. (\ref{entropy_D_0}) with the sign reversed:
\begin{align}\label{P_load_D}
\Bigl\langle P_{\mathrm{load}}^{\mathrm D}\Bigr\rangle=-\Bigl\langle P_{\mathrm{fric}}^{\mathrm D}\Bigr\rangle=-\left(\Gamma+\frac{\sigma^2}{G}\Biggl\langle\frac{\sin^2\theta}{V^2(\theta)}\Biggr\rangle_\theta\right)\left(\sum_{i=1}^N\biggl\langle\left(\omega_{\mathrm d}^{(i)}\right)^2\biggr\rangle\right)=-L_{11}^{-1}\left(\sum_{i=1}^N\biggl\langle\left(\omega_{\mathrm d}^{(i)}\right)^2\biggr\rangle\right),
\end{align}
which corresponds to the averaged brake power $\langle P_{\mathrm{rel}}\rangle$ due to the relative motion of the weakly-coupled Stirling engines, but this result is different from the approximation $\displaystyle\langle P_{\mathrm{rel}}\rangle\approx-L_{11}^{-1}\left(\sum_{i=1}^N\Bigl\langle\omega_{\mathrm d}^{(i)}\Bigr\rangle^2\right)$ given by Eq. (\ref{P_rel1}). The reason why the analysis based on the conceptual model has different results from the analysis using the quasilinear response relations (\ref{quasilinear}) will be discussed in subsection E. Another issue is the difference in energy dissipation due to viscous friction between the conceptual model and the original coupled engine model: the averaged energy dissipation rate due to viscous friction of the coupled Stirling engine is given by $\displaystyle\sum_{i=1}^N\Gamma\omega_i^2$, whereas in the conceptual model it is given by $\displaystyle\left(\Gamma+\frac{\sigma^2}{G}\Biggl\langle\frac{\sin^2\theta}{V^2(\theta)}\Biggr\rangle_\theta\right)\left[N\omega_{\mathrm m}^2+\sum_{i=1}^N\left(\omega_{\mathrm d}^{(i)}\right)^2\right]=\left(\Gamma+\frac{\sigma^2}{G}\Biggl\langle\frac{\sin^2\theta}{V^2(\theta)}\Biggr\rangle_\theta\right)\left(\sum_{i=1}^N\omega_i^2\right)$. This is due to the fact that the averaged energy dissipation rate $\displaystyle\frac{\sigma^2}{G}\Biggl\langle\frac{\sin^2\theta}{V^2(\theta)}\Biggr\rangle_\theta\left(\sum_{i=1}^N\omega_i^2\right)$ in the conceptual model comes from the heat leakage of the coupled Stirling engines (See Appendix B for details).

\subsection{Numerical experiments}
We conduct numerical experiments for five Stirling engines coupled in a chain structure with a uniform coupling strength of $K$ to confirm that synchronous and asynchronous transitions occur by changing the value of $K$, and that the averaged brake power and thermal efficiency increase or decrease due to these transitions. We also plot the results via the analysis using the quasilinear response relations (\ref{quasilinear}) and the conceptual model constructed in subsection B, and confirm the validity of the theory we have developed.

\begin{figure*}[htbp]
\begin{center}
\includegraphics[width=185mm]{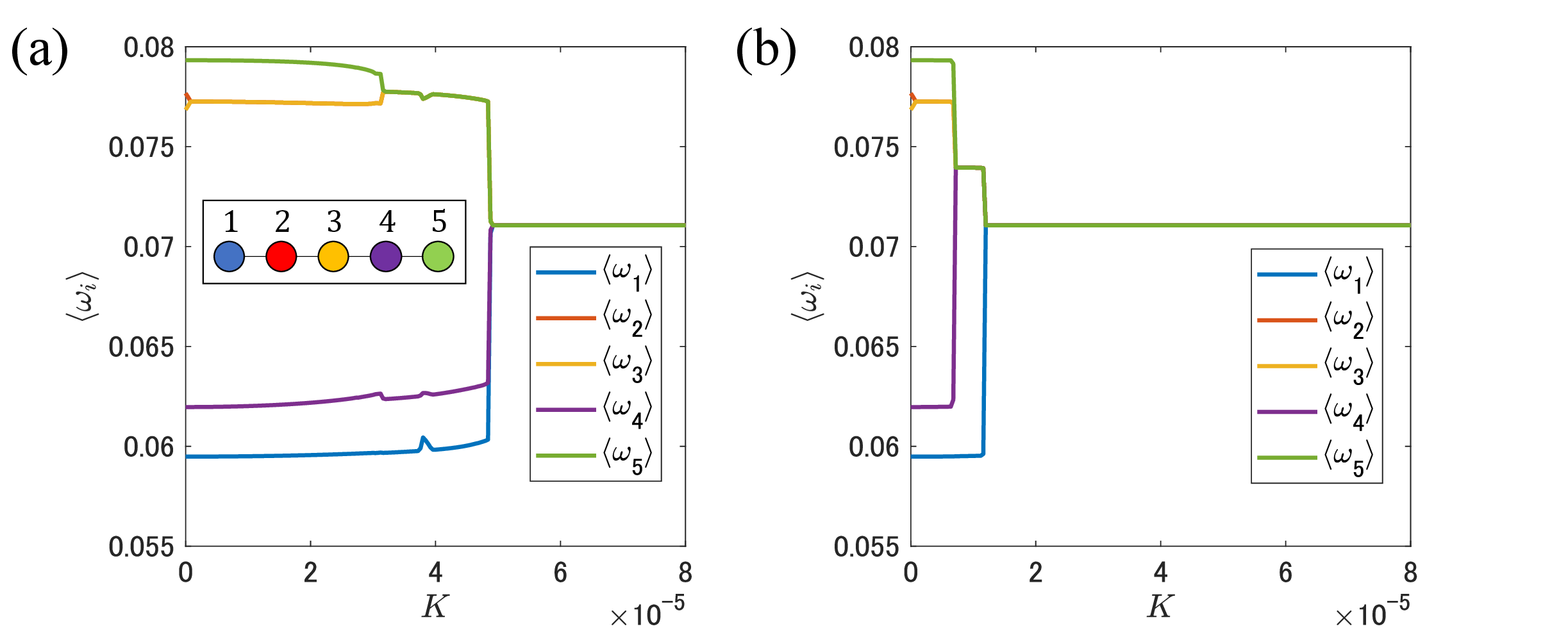}
\end{center}
\caption{Dependence relation between the effective frequency
$\langle\omega_i\rangle$ and the coupling strength $K$ of five engines coupled in a chain with a uniform coupling strength $K$ for (a) forward process and (b) backward process, where $T_{\rm load}^{(i)}$ are set as $T_{\rm load}^{(1)}=2.3038\times10^{-5}, T_{\rm load}^{(2)}=4.2662\times10^{-6}, T_{\rm load}^{(3)}=5.1195\times10^{-6}, T_{\rm load}^{(4)}=2.0478\times10^{-5}$, and $T_{\rm load}^{(5)}=2.5597\times10^{-6}$. Other parameters are the same as those used in Fig.\ref{assump_illu}. A schematic diagram of the five coupled engines is inserted in the middle of Fig. (a).
}
\label{5eng_syn_asyn}
\end{figure*}

Figure \ref{5eng_syn_asyn} shows the dependence of the coupling strength $K$ on the effective frequency $\langle\omega_i\rangle$ obtained from numerical experiments for the coupled system (\ref{dynamics0})-(\ref{dynamics}), where the forward (backward) process corresponds to the situation in which the value of $K$ is increased (decreased). The experimental results show that the ways in which transitions of $\langle\omega_i\rangle$ occur are different for the forward and backward processes, which is due to the coexistence of multiple stable states in the coupled system, and that partial synchronous (asynchronous) transitions occur sequentially in the process leading up to the complete synchronous (asynchronous) transitions.

\begin{figure*}[htbp]
\begin{center}
\includegraphics[width=170mm]{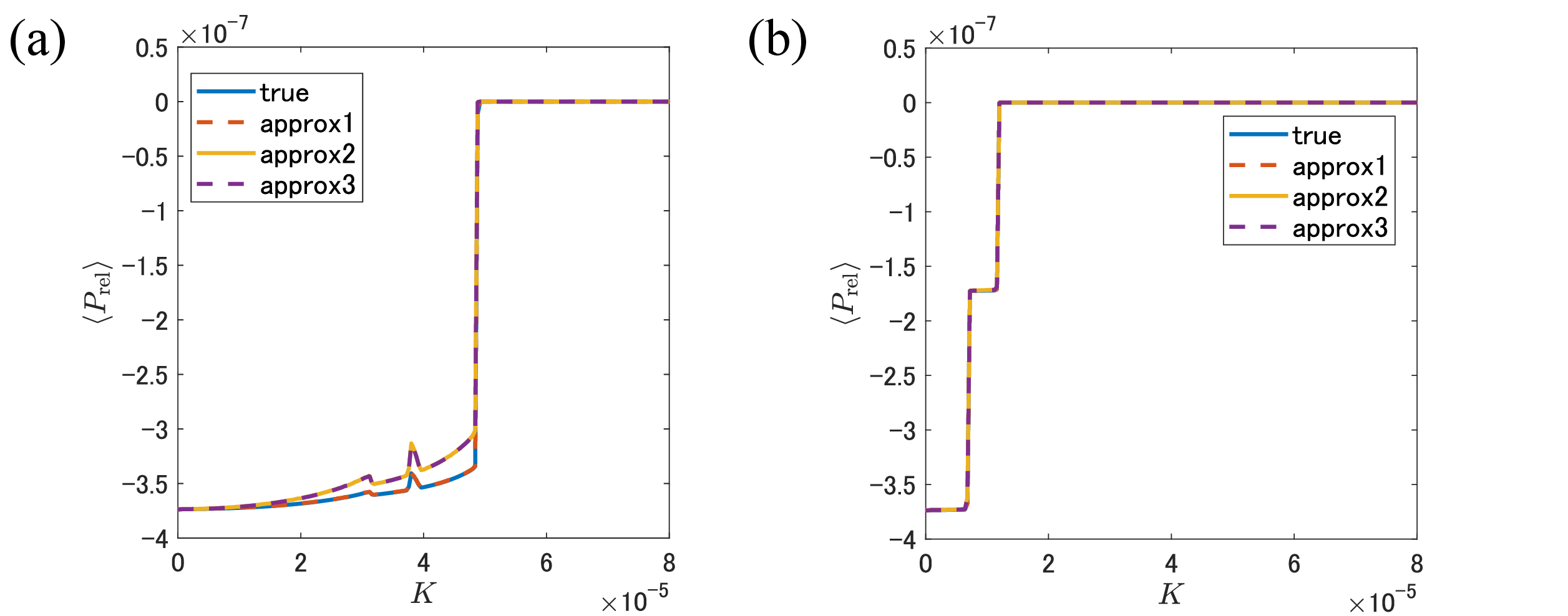}
\end{center}
\cprotect\caption{Dependence of the coupling strength $K$ on the averaged brake power due to relative motion $\langle P_{\mathrm{rel}}\rangle$, calculated for coupled system models (\ref{dynamics0})-(\ref{dynamics}). The true values (\verb|true|) were calculated from model (\ref{dynamics0})-(\ref{dynamics}), while the approximate values (\verb|approx1|, \verb|approx2| and \verb|approx3|) were calculated from Eqs. (\ref{P_rel}), (\ref{P_rel1_0}), and (\ref{P_rel1}), respectively. The values of $\Bigl\langle\omega_{\mathrm d}^{(i)}\Bigr\rangle$ and $\displaystyle\sum_{j\in\mathcal{N}_i}K_{ij}\Bigl\langle\sin\left(\theta_i-\theta_j\right)\Bigr\rangle$ contained in $\bigl\langle P_{\mathrm{rel}}\rangle$ were obtained from numerical experiments in the calculation of approximate values. (a): forward process, (b): backward process.
}\label{P_rel_realmodel}
\end{figure*}

\begin{figure*}[htbp]
\begin{center}
\includegraphics[width=170mm]{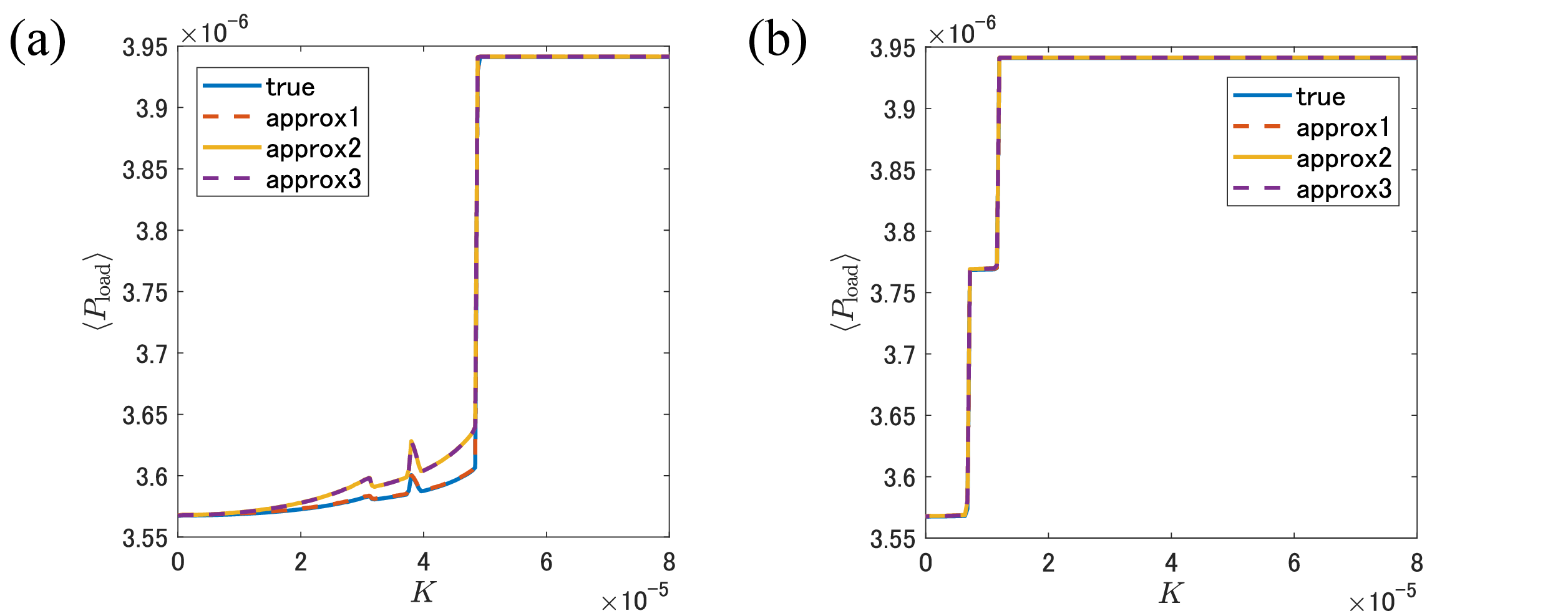}
\end{center}
\cprotect\caption{Dependence of the coupling strength $K$ on the averaged brake power $\langle P_{\mathrm{load}}\rangle$, calculated for coupled system models (\ref{dynamics0})-(\ref{dynamics}). The true values (\verb|true|) were calculated from model (\ref{dynamics0})-(\ref{dynamics}). The values of $\langle \omega_{\mathrm m}\rangle$ required for the calculation of the approximate values were computed using analytical solutions obtained from the quasilinear response relations (\ref{quasilinear}), while $\langle P_{\mathrm{rel}}\rangle$ was calculated in three different ways, using Eqs. (\ref{P_rel}) (corresponding to \verb|approx1|), (\ref{P_rel1_0}) (corresponding to \verb|approx2|) and (\ref{P_rel1}) (corresponding to \verb|approx3|). The values of $\Bigl\langle\omega_{\mathrm d}^{(i)}\Bigr\rangle$ and $\displaystyle\sum_{j\in\mathcal{N}_i}K_{ij}\Bigl\langle\sin\left(\theta_i-\theta_j\right)\Bigr\rangle$ contained in $\bigl\langle P_{\mathrm{rel}}\rangle$ were obtained from numerical experiments in the calculation of approximate values. (a): forward process, (b): backward process.
}\label{P_load_realmodel}
\end{figure*}

\begin{figure*}[htbp]
\begin{center}
\includegraphics[width=170mm]{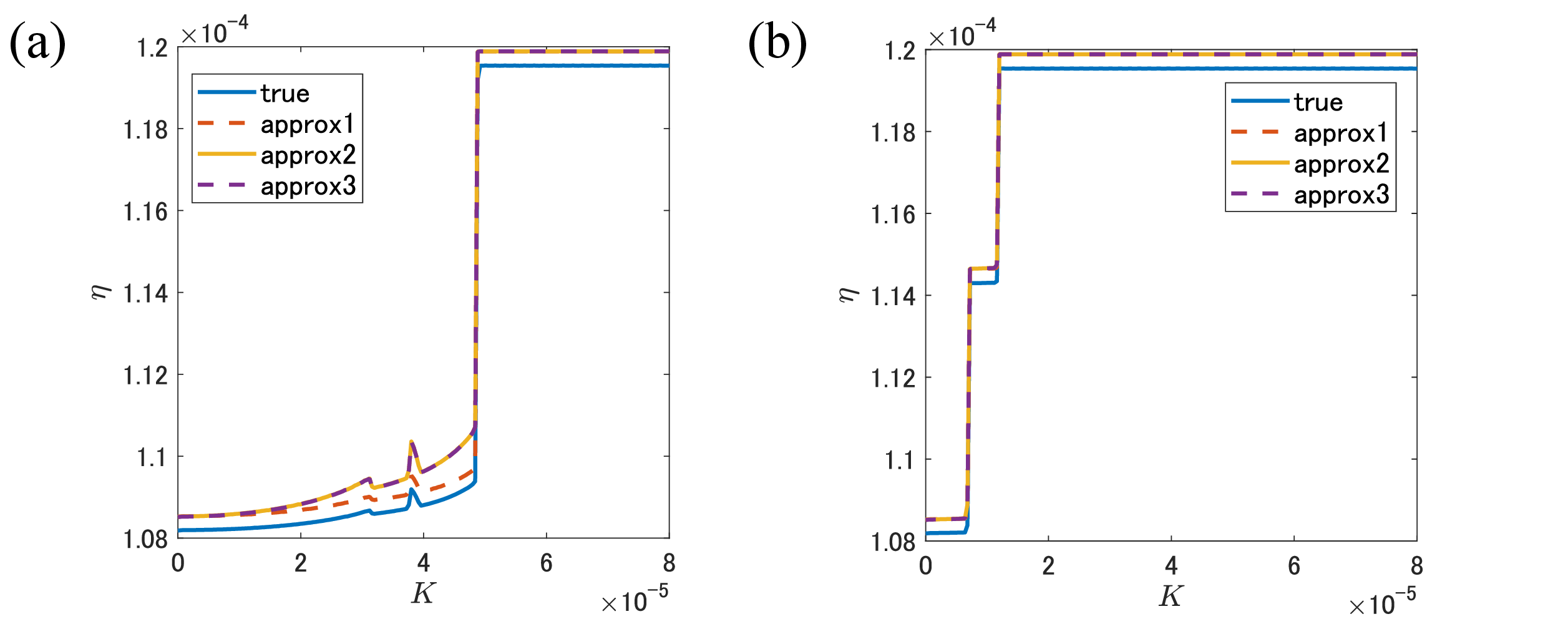}
\end{center}
\cprotect\caption{Dependence of the coupling strength $K$ on the thermal efficiency $\eta$, calculated for coupled system models (\ref{dynamics0}),(\ref{dynamics}) and (\ref{heatflux}). The true values (\verb|true|) were calculated from model (\ref{dynamics0}), (\ref{dynamics}) and (\ref{heatflux}). The values of $\langle \omega_{\mathrm m}\rangle$ and $\sum_{i=1}^N\Bigl\langle J_{Q_{\mathrm b}}^{(i)}\Bigr\rangle$ required for the calculation of the approximate values were computed using analytical solutions obtained from the quasilinear response relations (\ref{quasilinear}), while $\langle P_{\mathrm{rel}}\rangle$ was calculated in three different ways, using Eqs. (\ref{P_rel}) (corresponding to \verb|approx1|), (\ref{P_rel1_0}) (corresponding to \verb|approx2|) and (\ref{P_rel1}) (corresponding to \verb|approx3|). The values of $\Bigl\langle\omega_{\mathrm d}^{(i)}\Bigr\rangle$ and $\displaystyle\sum_{j\in\mathcal{N}_i}K_{ij}\Bigl\langle\sin\left(\theta_i-\theta_j\right)\Bigr\rangle$ contained in $\bigl\langle P_{\mathrm{rel}}\rangle$ were obtained from numerical experiments in the calculation of approximate values. (a): forward process, (b): backward process.
}\label{eta_realmodel}
\end{figure*}

Figure \ref{P_rel_realmodel}, \ref{P_load_realmodel} and \ref{eta_realmodel} plot the averaged brake power owing to the relative motion $\langle P_{\mathrm{rel}}\rangle$, the averaged brake power $\langle P_{\mathrm{load}}\rangle$, and the thermal efficiency $\eta$ calculated for the coupled system (\ref{dynamics0}), (\ref{dynamics}) and (\ref{heatflux}). Here, the true values (\verb|true|) were calculated directly from model (\ref{dynamics0}), (\ref{dynamics}) and (\ref{heatflux}), while the approximate values (\verb|approx1|, \verb|approx2| and \verb|approx3|) were calculated from the quasilinear response relations (\ref{quasilinear}) and approximate formula (\ref{sum_P_coupling}). Specifically, the values of $\langle \omega_{\mathrm m}\rangle$ and $\sum_{i=1}^N\Bigl\langle J_{Q_{\mathrm b}}^{(i)}\Bigr\rangle$ required for the calculation of the approximate values above were computed using analytical solutions obtained from the quasilinear response relations (\ref{quasilinear}), while $\langle P_{\mathrm{rel}}\rangle$ was calculated in three different ways, using Eqs. (\ref{P_rel}) (corresponding to \verb|approx1|), (\ref{P_rel1_0}) (corresponding to \verb|approx2|) and (\ref{P_rel1}) (corresponding to \verb|approx3|). Among the three approximations for $\langle P_{\mathrm{rel}}\rangle$, Eq. (\ref{P_rel1}) is essentially the most important one since it can explain the optimality of the synchronous state. However, to compare the errors induced by these three approximations, plots of formulas using Eqs. (\ref{P_rel}) and (\ref{P_rel1_0}) are also included in the same figure. Since the values of $\Bigl\langle\omega_{\mathrm d}^{(i)}\Bigr\rangle$ and $\sum_{j\in\mathcal{N}_i}K_{ij}\Bigl\langle\sin\left(\theta_i-\theta_j\right)\Bigr\rangle$ contained in $\bigl\langle P_{\mathrm{rel}}\rangle$ are difficult to obtain analytically, we used numerical values of them in the calculation of approximate values. The experimental results show that the approximations agree with the true values over a wide range of parameters, and that synchronous (asynchronous) transition increases (decreases) the values of $\langle P_{\mathrm{rel}}\rangle$, $\langle P_{\mathrm{load}}\rangle$ and $\eta$. In regions where the coupling strength is weak enough, the three approximations above have nearly identical deviations from the true value, while the deviations from the true values of \verb|approx2| and \verb|approx3| become slightly larger when the coupling strength is further increased, and they deviate from the true value to the same degree as \verb|approx1| again after the synchronous transition occurs.

\begin{figure*}[htbp]
\begin{center}
\includegraphics[width=170mm]{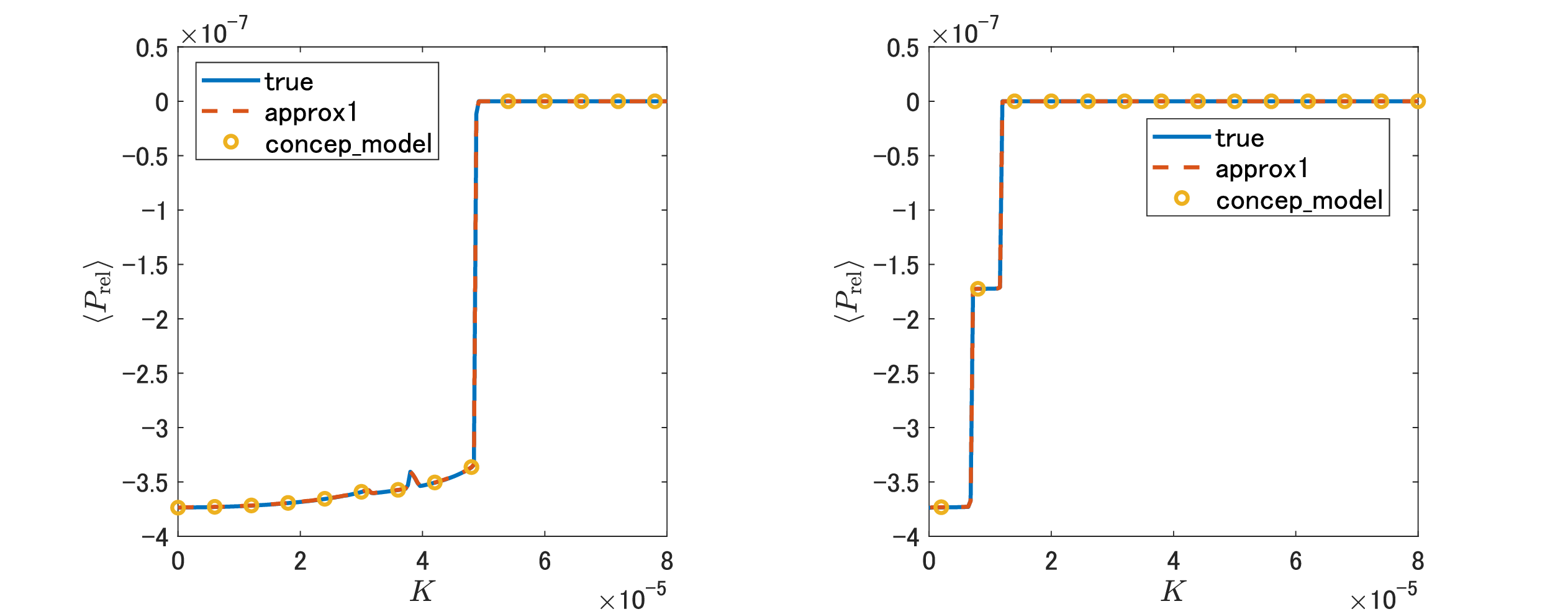}
\end{center}
\cprotect\caption{Dependence of the averaged brake power due to relative motion $\langle P_{\mathrm{rel}}\rangle$ on the coupling strength $K$, calculated through the conceptual model (corresponding to \verb|concep_model|) constructed in subsection B. The true values (\verb|true|) and approximate values \verb|approx1| calculated for the original coupled system models (\ref{dynamics0})-(\ref{dynamics}) are also added for comparison. Here, the averaged brake power in isolated system D was calculated using Eq. (\ref{entropy_D_0}) with the sign reversed. (a): forward process, (b): backward process.
}\label{P_rel_approxmodel}
\end{figure*}

\begin{figure*}[htbp]
\begin{center}
\includegraphics[width=170mm]{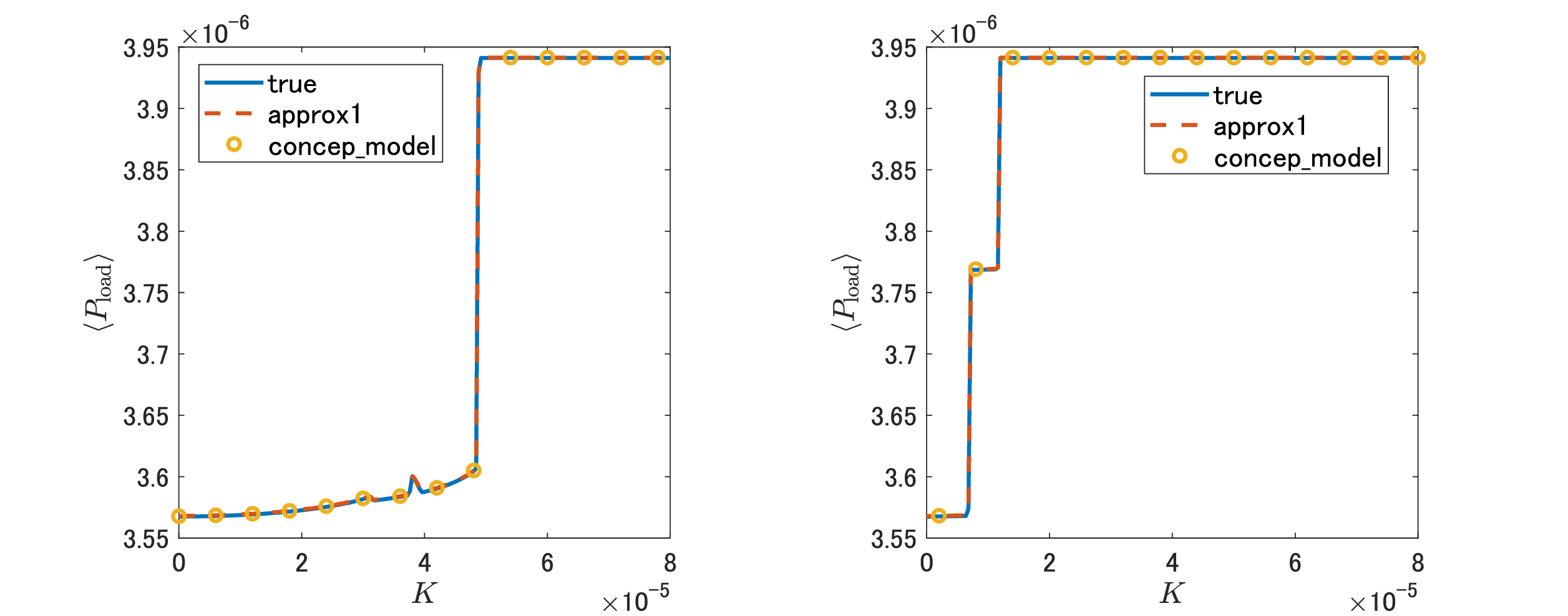}
\end{center}
\cprotect\caption{Dependence of the averaged brake power $\langle P_{\mathrm{load}}\rangle$ on the coupling strength $K$, calculated through the conceptual model (corresponding to \verb|concep_model|) constructed in subsection B. The true values (\verb|true|) and approximate values \verb|approx1| calculated for the original coupled system models (\ref{dynamics0})-(\ref{dynamics}) are also added for comparison. Here, the averaged brake power in isolated system D was calculated using Eq. (\ref{entropy_D_0}) with the sign reversed. (a): forward process, (b): backward process.
}\label{P_load_approxmodel}
\end{figure*}

\begin{figure*}[htbp]
\begin{center}
\includegraphics[width=170mm]{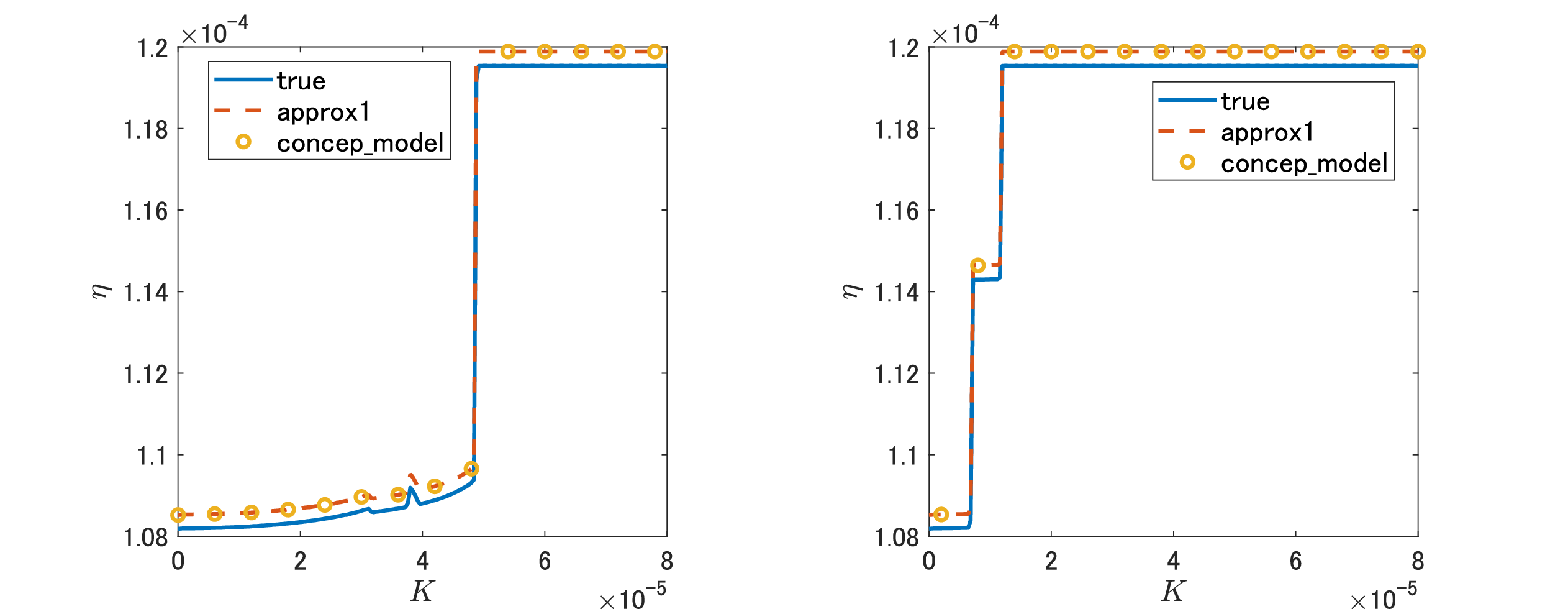}
\end{center}
\cprotect\caption{Dependence of the thermal efficiency $\eta$ on the coupling strength $K$, calculated through the conceptual model (corresponding to \verb|concep_model|) constructed in subsection B. The true values (\verb|true|) and approximate values \verb|approx1| calculated for the original coupled system models (\ref{dynamics0}), (\ref{dynamics}), and (\ref{heatflux}) are also added for comparison. Here, the averaged brake power in isolated system D was calculated using Eq. (\ref{entropy_D_0}) with the sign reversed. (a): forward process, (b): backward process.
}\label{eta_approxmodel}
\end{figure*}

Figures \ref{P_rel_approxmodel}, \ref{P_load_approxmodel} and \ref{eta_approxmodel} show the values of $\langle P_{\mathrm{rel}}\rangle$, $\langle P_{\mathrm{load}}\rangle$ and $\eta$, respectively, calculated through the conceptual model (corresponding to \verb|concep_model|) constructed in subsection B. The true values (\verb|true|) and approximate values \verb|approx1| calculated for the original coupled system models (\ref{dynamics0}), (\ref{dynamics}) and (\ref{heatflux}) are also added for comparison. Here, the averaged brake power in isolated system D was calculated using Eq. (\ref{entropy_D_0}) with the sign reversed (Note that the averaged entropy production rate is numerically equal to the sign-reversed averaged brake power in system D, since the temperature of the heat reservoir in system D is approximated to 1). It can be confirmed that the values calculated from the conceptual model almost overlap with \verb|approx1|, which means that large deviations from the true values as those confirmed in \verb|approx2| and \verb|approx3| have not occured for the conceptual model, and are thus a good approximation even near the synchronous transition point.


\subsection{Discussion}
As described in subsection B, the averaged power (\ref{P_load_D}) in the constructed subsystem D differs from the approximation (\ref{P_rel1}) of $\langle P_{\mathrm{rel}}\rangle$ obtained in subsection A. This difference is also reflected in the numerical experiments in subsection D. Indeed, it can be confirmed from the numerical experiments that the approximations using Eqs. (\ref{P_rel1_0}) and (\ref{P_rel1}) deviate slightly from the true value in the region just before the synchronous transition, while the analysis using the conceptual model does not show the above phenomenon. In this subsection, we consider the reasons for this and review the theory we have developed so far.

Before evaluating the approximation error by Eqs. (\ref{P_rel1_0}) and (\ref{P_rel1}), we first review the approximation procedure used to obtain the quasilinear response relations (\ref{quasilinear}) (c.f. Appendix A). We approximated $\omega_i$ to a certain constant $\overline{\omega_k^{(i)}}$ each time the engine makes one rotation there since the variation of $\omega_i$ during one rotation of the phase $\theta_i$ is sufficiently small. Therefore, the approximation error that occurs when calculating the long-time average of the driving force portion of the engine or the heat flux is of the same order of magnitude as the time average of the minute approximation error that occurs for each rotation of the engine. In this sense, the quasilinear response relations (\ref{quasilinear}) are mathematically valid relations. On the other hand, Eqs. (\ref{P_rel1_0}) and (\ref{P_rel1}) are calculated via Eq. (\ref{sum_P_coupling}), which is obtained by approximating the value of $\omega_i$ to its long-time average $\langle\omega_i\rangle$, so the approximation accuracy is worse than that of relations (\ref{quasilinear}). If the coupling strength is small enough, the value of $\omega_i$ at any given time fluctuates little from its long time average $\langle\omega_i\rangle$, so $\langle\omega_i\rangle$ is not much different from the time average of one rotation of $\theta_i$, but if the coupling strength is increased to some extent, the range of variation of $\omega_i$ on the long time scale becomes larger, and $\omega_i$ can deviate significantly from its long time average $\langle\omega_i\rangle$. Therefore, as confirmed by numerical experiments, the approximation errors of Eqs. (\ref{P_rel1_0}) and (\ref{P_rel1}) are larger than those of (\ref{P_rel}) in the region just before the synchronous transition occurs. After the synchronous transition occurs, the right-hand side of Eq.(\ref{sum_P_coupling}) becomes exactly zero and the approximation error due to the above approximation disappears.

The above is a discussion of the approximation error of the analysis based on the quasilinear response relations (\ref{quasilinear}) from a mathematical point of view. To give a discussion from the point of view of nonequilibrium thermodynamics, let us look back again at the calculation of the entropy production rate (\ref{entropy_first})-(\ref{entropy_final}) in subsection A. The intermediate formula (\ref{entropy_approx}) uses approximation (\ref{P_K_approx}), which is essentially the same as approximation (\ref{sum_P_coupling}). Since the averaged power exerted by the coupling forces between engines on the total coupled system is zero, the averaged entropy production rate for the total system should normally be expressed without using the long-time average of the coupling force. That is, the real thermodynamic force conjugate to $\Bigl\langle\omega_{\mathrm d}^{(i)}\Bigr\rangle$ should be the one expressing the load heterogeneity, $\displaystyle\frac{1}{N}\left(\sum_{k=1}^{N}T_{\mathrm{load}}^{(k)}\right)-T_{\mathrm{load}}^{(i)}$, not $\displaystyle\frac{1}{N}\left(\sum_{k=1}^{N}T_{\mathrm{load}}^{(k)}\right)-T_{\mathrm{load}}^{(i)}-\sum_{j\in\mathcal{N}_i}K_{ij}\Bigl\langle\sin\left(\theta_i-\theta_j\right)\Bigr\rangle$ exactly, and the coupling strength should be included in the response coefficients. This implies that the coupling strength is an internal parameter of the coupled system that regulates the fluxes $\Bigl\langle\omega_{\mathrm d}^{(i)}\Bigr\rangle$ caused by the load heterogeneity. The stronger the coupling strength is, the greater the suppression of fluxes $\Bigl\langle\omega_{\mathrm d}^{(i)}\Bigr\rangle$ due to the load heterogeneity is. In this case, however, it is difficult to obtain the response coefficients analytically (in fact, the relations between the fluxes and forces are not even linear, and show hysteresis structures that originate from the dependence on the state variables of the coupled system), so we dared to use the approximation (\ref{P_K_approx}) and included the averaged coupling force $\displaystyle-\sum_{j\in\mathcal{N}_i}K_{ij}\Bigl\langle\sin\left(\theta_i-\theta_j\right)\Bigr\rangle$ in the thermodynamic forces conjugate to the fluxes $\Bigl\langle\omega_{\mathrm d}^{(i)}\Bigr\rangle$ when calculating the averaged entropy production rate. On the other hand, we included the term $\displaystyle\sum_{i=1}^{N}\sum_{j\in\mathcal{N}_i}K_{ij}\Bigl\langle\sin\left(\theta_i-\theta_j\right)\Bigr\rangle\Bigl\langle\omega_{\rm d}^{(i)}\Bigr\rangle$ when computing $\bigl\langle P_{\rm rel}\bigr\rangle$ in Eq. (\ref{P_rel1_0}), which corresponds to the above operation of putting $\displaystyle-\sum_{j\in\mathcal{N}_i}K_{ij}\Bigl\langle\sin\left(\theta_i-\theta_j\right)\Bigr\rangle$ in the force conjugate to $\Bigl\langle\omega_{\mathrm d}^{(i)}\Bigr\rangle$. Thus, the inaccuracy in the analysis of the averaged brake power and the thermal efficiency using quasilinear response relations (\ref{quasilinear}) in subsection A is due to the fact that the coupling strength, which is essentially an internal parameter that regulates the fluxes due to load heterogeneity, was put into the thermodynamic forces that produce irreversibility. In other words, the discrepancy between the analytical results and the true values is due to the fact that the quasilinear response relations (\ref{quasilinear}) do not preserve the thermodynamic irreversibility of the original coupled system.

Let us now discuss the approximation errors that arise in the analysis using the conceptual model constructed in subsection B. The analysis of the averaged power and thermal efficiency for this conceptual model does not go through any approximate calculations, so the results are completely accurate if we assume that the model itself is valid. The conceptual model was constructed to satisfy the quasilinear response relations (\ref{quasilinear}), but since the relations include the terms of the averaged coupling forces, which depend on the dynamics of the original coupled engines, we cannot conclude directly that the conceptual model satisfies the quasilinear response relations (\ref{quasilinear}) unless the terms of the averaged coupling forces in the conceptual model are in close agreement with those in the original coupled engine model. As shown in subsection C, the terms of the averaged coupling forces in the conceptual model and the original coupled model are nearly identical given that conditions (\ref{quasilinear_N_single}), (\ref{condition1_nuturalfrequency}) and (\ref{condition2_nuturalfrequency}) are satisfied, so the conceptual model satisfies the quasilinear response relations (\ref{quasilinear}) with good approximation accuracy. On the other hand, since the quasilinear response relations (\ref{quasilinear}) do not preserve the thermodynamic irreversibility of the original coupled system, a concern arises that the conceptual model constructed from the quasilinear response relations may not preserve the thermodynamic irreversibility as well. However, the averaged entropy production rate of the constructed isolated system D is given by Eq. (\ref{entropy_D_1}), which is not consistent with that determined from the quasilinear response relations (\ref{linear_D}). The approximate formula (\ref{approx_0}), which is practically equivalent to the approximate formula (\ref{sum_P_coupling}), was used in the calculation of the averaged entropy production rate for the isolated system D to ensure that system D satisfies the quasilinear response relations (\ref{linear_D}). It can be seen from the averaged entropy production rate (\ref{entropy_D_1}) for the isolated system D that the thermodynamic fluxes and forces of the conceptual model are in perfect agreement with the real thermodynamic fluxes and forces of the original coupled system if the averaging error is ignored, and thus, unlike the quasilinear response relations (\ref{quasilinear}), the conceptual model constructed in subsection B correctly reflects the nature of the thermodynamic irreversibility of the original coupled system. This is the reason why the analytical results using the quasilinear response relations (\ref{quasilinear}) differ from those using the conceptual model. Therefore, the conceptual model provides accurate analytical results for the averaged power and thermal efficiency without requiring the unknown response relations between the real thermodynamic fluxes and forces.

\section{Analysis and comparison of rotational dynamics between original and conceptual models for two-engine systems}
In Section III, a conceptual model was constructed for weakly-coupled LTD Stirling engines. However, the difference of the behavior between that conceptual model and the original one is not obvious. In this section, we focus on the synchronous and asynchronous transitions of the differential equations (\ref{dynamics0})-(\ref{dynamics}) describing the rotational motion of the weakly coupled Stirling engines and the corresponding conceptual model obtained in Section III for a two-coupled engine system, and discuss the differences between the two in their long-time behavior.

\begin{figure*}[htbp]
\begin{center}
\includegraphics[width=185mm]{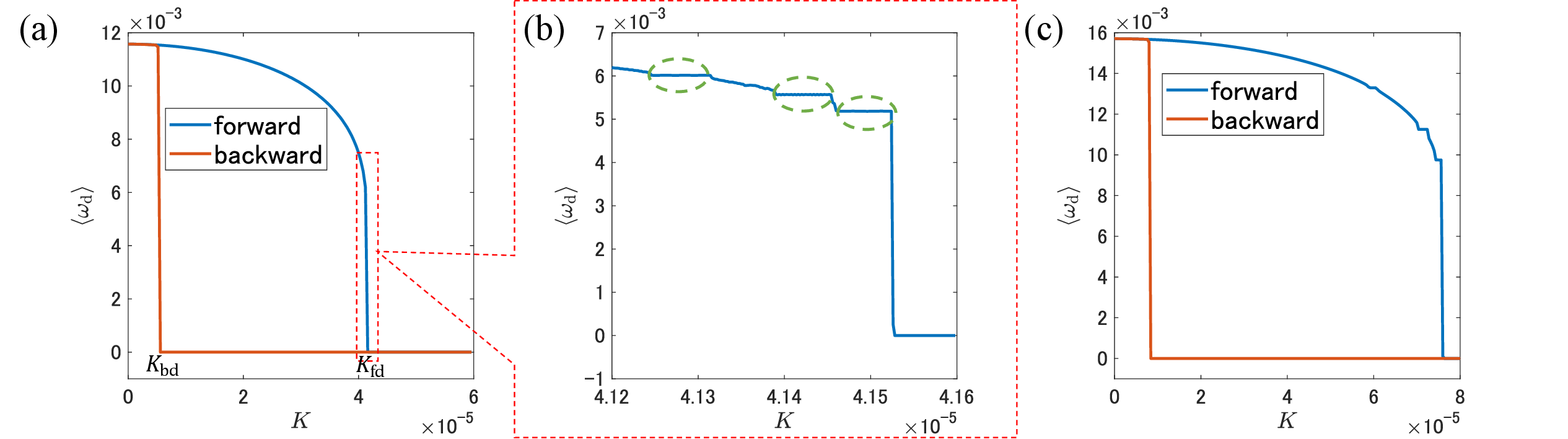}
\end{center}
\caption{(a). Dependence relation between the effective frequency difference 
$\langle\omega_{\rm d}\rangle$
and the coupling strength $K$. All the parameters are set to the same values as those used in Fig. (\ref{assump_illu}).  (b) Enlarged view of $\langle\omega_{\rm d}\rangle$ near the synchronous transition point. It can be seen that $\langle\omega_{\rm d}\rangle$ is falling in a stair-step manner. Nearly flat regions, some of which are enclosed by green dashed circles, exhibit $m:n$ synchronous states, while other areas exhibit quasi-periodic states. There exist many such nearly flat regions other than the areas enclosed by the green dashed lines, such as the areas between these enclosed areas. (c). Dependence relation between the effective frequency difference 
$\langle\omega_{\rm d}\rangle$
and the coupling strength $K$, where the value of $T_{\rm load}^{(2)}$ is changed to $T_{\rm load}^{(2)}=1.7065\times10^{-5}$.
}
\label{demonstair}
\end{figure*}

We first consider the dynamics before averaging the fast variables, where the equation of motion of engine $i$ and $j$ ($i,j\in\{1,2\}, i\neq j$) is given by
\begin{subequations}
\begin{equation}\label{2_dynamics0}
\frac{d\theta_i}{dt}=\omega_i,
\end{equation}
\begin{equation}\label{2_dynamics}
\frac{d\omega_i}{dt}=\sigma\left(\frac{T(\theta_i,\omega_i)}{V(\theta_i)}-P_{\rm air}\right)\sin\theta_i-\Gamma\omega_i-T_{\rm load}^{(i)}-K\sin(\theta_i-\theta_j).
\end{equation}
\end{subequations}
The natural frequency of each engine can be adjusted by changing the load torque acting on the crank. Here we set all the parameters to the same values as those used in Fig. \ref{assump_illu}. Figure \ref{demonstair}(a) shows the dependence relation between $\langle\omega_{\rm d}\rangle\equiv\langle\omega_1\rangle-\langle\omega_2\rangle$ and $K$, where the forward (backward) process corresponds to the situation in which the value of $K$ is increased (decreased). The hysteresis structure is due to the coexistence of a stable synchronous state and asynchronous states \cite{PhysRevResearch.5.043268}: only asynchronous motions that appear to be quasi-periodic trajectories exist until $K$ exceeds a certain threshold $K_{\rm bd}$, where a stable limit cycle corresponding to a stable synchronous state and a saddle limit cycle corresponding to an unstable synchronous state emerge due to a saddle-node bifurcation; as $K$ approaches $K_{\rm fd}$, asynchronous trajectories approach the saddle limit cycle, which makes $\langle\omega_{\rm d}\rangle$ approach 0; when $K$ exceeds $K_{\rm fd}$, asynchronous trajectories disappear and the stable synchronous state becomes globally stable (See \cite{PhysRevResearch.5.043268} for details). This is similar to the phenomenon seen in homoclinic bifurcations \cite{Strogatz}.

Although the synchronous transition exhibits the characteristics of a homoclinic bifurcation, we cannot conclude directly from the above analysis that the synchronous transition is caused by a homoclinic bifurcation. Therefore, we will analyze the process leading to the synchronous transition in more detail. If one-step increments of the coupling strength near $K_{\rm fd}$ are made finer, a phenomenon called the devil's staircase \cite{Pikovsky}, in which quasi-periodic and $m:n$ synchronous states $(m\neq n)$ appear alternately as the coupling strength is increased, can be observed. (Fig. \ref{demonstair}(b)). The areas where $\langle\omega_{\rm d}\rangle$ is nearly flat (e.g. the areas that are enclosed by green dashed circles) exhibit $m:n$ synchronous states, which are limit cycles circling around the phase space $\mathbb{T}^2\times\mathbb{R}^2$, while the rest exhibit quasi-periodic states. Note that there exist many such nearly flat regions other than the areas enclosed by the green dashed lines, such as the areas between these enclosed areas. Like the $1:1$ synchronous states, a pair of $m:n$ synchronous states, one is stable and the other is unstable, are generated by a saddle-node bifurcation when the coupling strength $K$ is increased up to a certain threshold $K_{m:n}^1$. If we further increase the value of $K$, the flat part of the staircase collapses, suggesting that the above two $m:n$ synchronized states disappear due to a saddle-node bifurcation that occurs at another threshold $K_{m:n}^2$. The disappearance of the stable $m:n$ synchronous state causes the system in that state to either move to a quasi-periodic state or converge to another $m:n$ synchronous state. It should be noted that before the transition to a stable $m:n$ synchronous state occurs, that $m:n$ synchronous state has already been created by a saddle-node bifurcation. Accordingly, the flat parts of the devil's staircase are narrower than the regions where the actual $m:n$ synchronous states can exist. Since Fig. \ref{demonstair}(b) suggests that the $1:1$ synchronous transition is a discontinuous transition caused by the disappearance of a certain stable $m:n$ synchronous state at a saddle-node bifurcation, the synchronous transition is not strictly a homoclinic bifurcation. To be precise, the process leading to the $1:1$ synchronous transition is the one in which the transitions of an asynchronous state (quasi-periodic or an $m:n$ synchronous state) to another $m:n$ synchronous state and the collapse of $m:n$ synchronous states due to saddle node bifurcations occur alternately while reserving the feature of a homoclinic bifurcation in that the asynchronous states gradually approach the saddle limit cycle corresponding to the unstable $1:1$ synchronous state. However, if $\Delta\omega_{\rm n}$ is sufficiently small compared to $\omega_{\rm n}^{(i)}$, the $m:n$ synchronous regions turn to be sufficiently narrow compared to the range of the variation of the coupling strength $K$ up to $K_{\rm fd}$, so on a macroscopic scale, the synchronous transition can be considered to be caused by a homoclinic bifurcation approximately. If the load torques are adjusted so that $\Delta\omega_{\rm n}$ is made larger, the nearly flat $m:n$ synchronous regions in the devil's staircase become wider, and the coarse-grained homoclinic bifurcation structure in the synchronous transition breaks down. Figure \ref{demonstair}(c) shows the dependence relation between $\langle\omega_{\rm d}\rangle$ and $K$ with the value of $\Delta\omega_{\mathrm n}$ slightly larger than that in Fig. \ref{demonstair}(a). The areas corresponding to $m:n$ synchronous states are wider than those observed in Fig. \ref{demonstair}(a), and it can be seen even on a macroscopic scale that the synchronous transition occurs discontinuously. Furthermore, the $m:n$ synchronous regions observed in the devil's staircase tend to become wider as the synchronous transition point is approached in the forward process. This dependence of the width of the $m:n$ synchronous regions on $\Delta\omega_{\mathrm n}$ and $K$ can be explained by the distorted V-shaped structure of Arnold's tongues \cite{Pikovsky}.

Next, we consider the differential equations of the conceptual model (i.e., the averaged differential equations w.r.t. Eqs. (\ref{dynamics0}) and (\ref{dynamics})). The equations describing the rotational motion of subsystem D are given by
\begin{subequations}
\begin{equation}\label{diff0}
\frac{d\theta_{\rm d}}{dt}=\omega_{\rm d},
\end{equation}
\begin{equation}\label{diff}
\frac{d\omega_{\rm d}}{dt}+\left(\Gamma+\frac{\sigma^2}{G}\Biggl\langle\frac{\sin^2\theta}{V^2(\theta)}\Biggr\rangle_\theta\right)\omega_{\rm d}=-\left(T_{\rm load}^{(1)}-T_{\rm load}^{(2)}\right)-2K\sin\theta_{\rm d}.
\end{equation}
\end{subequations}
Equations (\ref{diff0}) and (\ref{diff}) have the same form as those of a damped nonlinear pendulum that is driven by a state-independent force \cite{Strogatz}, or the phase difference of a two-node power grid consisting of one generator and one consumer \cite{manik2014supply,rohden2012self}, where the equation of motion is given by
\begin{equation}\label{powergrid}
\frac{d^2x}{dt^2}=P_0-\alpha\frac{dx}{dt}+2K\sin x.
\end{equation}
The dynamic behavior of the differential equation given by Eq. (\ref{powergrid}) has already been well investigated \cite{Strogatz,manik2014supply,rohden2012self}. For $K<P_0/2$, no fixed point exists and the phase difference approaches a stable limit cycle circling the phase space $\mathbb{T}\times\mathbb{R}$. For $K>P_0/2$, there is a stable fixed point and a saddle point, which are created by a saddle-node bifurcation at $K=P_0/2$. The stable limit cycle coexists with the fixed points if the increase in kinetic energy due to $P_0$ compensates the decrease due to friction. Thus, the parameter space of the system can be divided into three regions: a region where a limit cycle is globally stable, a region where a fixed point is globally stable, and a region where both coexist. At the boundary that separates the globally stable fixed point regime and the
coexistence regime, a homoclinic orbit of the saddle fixed point emerges from the limit cycle in a homoclinic bifurcation. This boundary can be obtained in the low-friction limit, which is given by $\displaystyle P_0\approx\frac{4\sqrt{2}}{\pi}\cdot\alpha\sqrt{K}$ and is found to be consistent with the actual value when $\alpha/\sqrt{K}<0.6$. The homoclinic boundary intersects the saddle-node bifurcation line ($K=P_0/2$) at a numerically determined value of $\alpha/\sqrt{K}\approx1.69$. For $\alpha/\sqrt{K}\gtrsim1.69$, the saddle-node bifurcation and the homoclinic bifurcation combine to a saddle-node homoclinic bifurcation.

\begin{figure*}[htbp]
\begin{center}
\includegraphics[width=100mm]{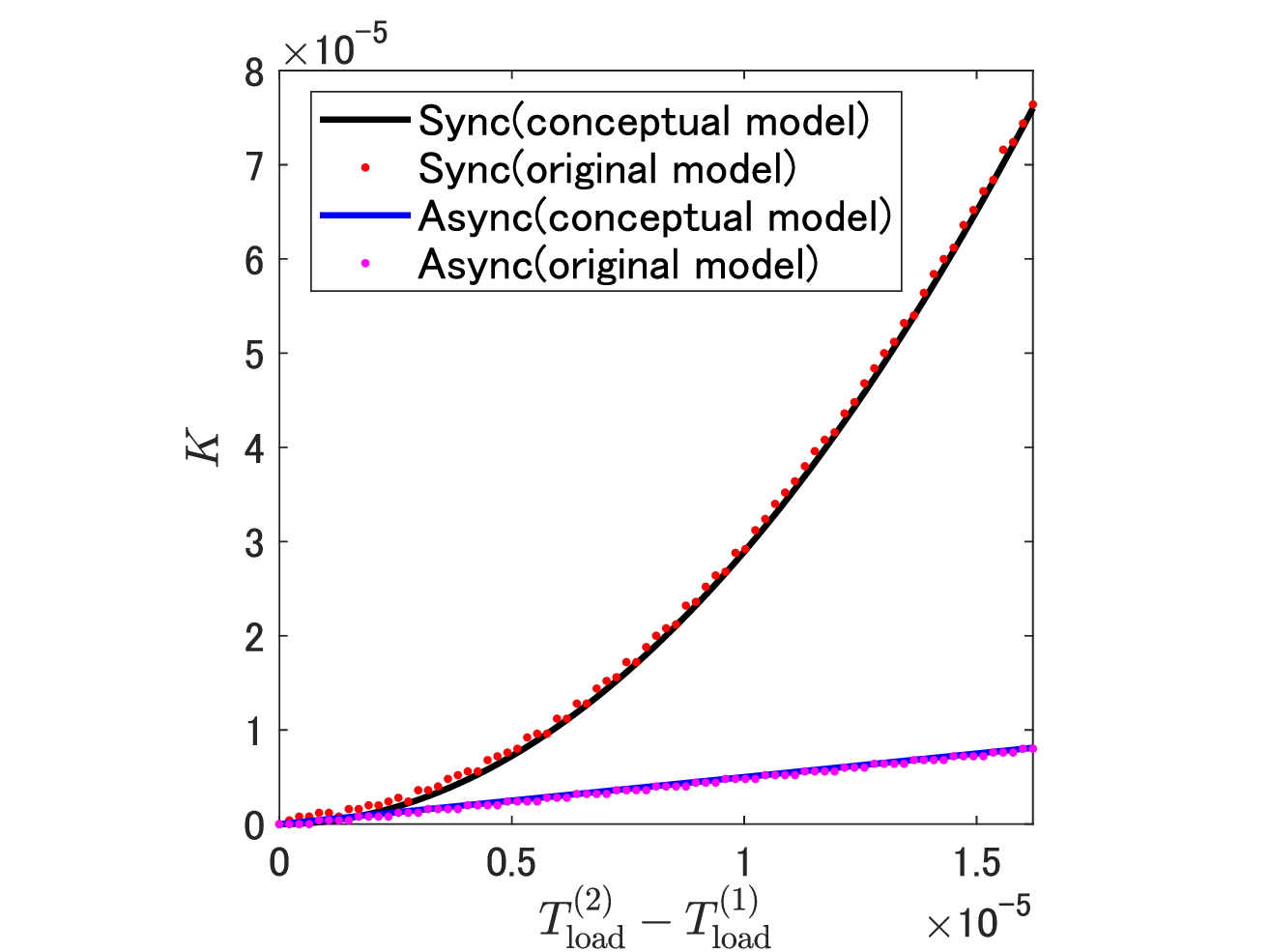}
\end{center}
\caption{Synchronous and asynchronous transition points of the averaged differential equations (\ref{diff0}) and (\ref{diff}) (solid lines) calculated using Eqs. (\ref{SD}) and (\ref{HOMO}), and of the original coupled system (\ref{2_dynamics0}) and (\ref{2_dynamics}) (discrete data) obtained from numerical experiments.
}\label{orin_ave}
\end{figure*}

Now we apply the above results to Eqs. (\ref{diff0}) and (\ref{diff}). The saddle-node bifurcation point $K_{\rm s}$ is obtained as
\begin{equation}\label{SD}
K_{\rm s}=\frac{1}{2}\left(T_{\rm load}^{(2)}-T_{\rm load}^{(1)}\right), 
\end{equation}
and the homoclinic bifurcation point $K_{\rm h}$ can be approximated by 
\begin{equation}\label{HOMO}
K_{\rm h}\approx\frac{\pi^2\left(T_{\rm load}^{(2)}-T_{\rm load}^{(1)}\right)^2}{32\left(\Gamma+\frac{\sigma^2}{G}\Bigl\langle\frac{\sin^2\theta}{V^2(\theta)}\Bigr\rangle_\theta\right)^2}
\end{equation}
given that $\displaystyle\frac{\Gamma+\frac{\sigma^2}{G}\Bigl\langle\frac{\sin^2\theta}{V^2(\theta)}\Bigr\rangle_\theta}{\sqrt{K_{\rm h}}}<0.6$. Figure \ref{orin_ave} plots the bifurcation points of the averaged differential equations (\ref{diff0}) and (\ref{diff}) calculated using Eqs. (\ref{SD}) and (\ref{HOMO}), and the synchronous and asynchronous transition points of the original coupled system (\ref{dynamics0}) and (\ref{dynamics}) obtained from numerical experiments. The excellent agreement of the values obtained from the averaged equations with the values obtained from the original model can be confirmed over a wide parameter range, which supports the validity of the averaged equations.

Comparing the original model (\ref{2_dynamics0}) and (\ref{2_dynamics}) and the averaged differential equation (\ref{diff0}) and (\ref{diff}), we find that the limit cycles corresponding to the stable and unstable $1:1$ synchronized states become stable and unstable fixed points, while limit cycles corresponding to $m:n$ synchronized states disappear through averaging, which implies that the devil's staircase that appeared leading up to the synchronous transitions can not be observed for the conceptual model. This suggests that the model obtained by averaging fast variables preserves the macroscopic structure but loses the detailed structure of the original model. The loss of the detailed structure is the result of the fact that the dynamics of the differential system is completely separated from the 4-dimensional coupled system through the averaging method.

\section{Summary and outlook}
This study has presented a theory of quasilinear irreversible thermodynamics for multiple weakly-coupled LTD Stirling engines with synchronous and asynchronous transitions. We analyzed the effects of synchronous and asynchronous transitions on power and thermal efficiency using the quasilinear response relations between formally defined thermodynamic fluxes and forces, and constructed a conceptual model satisfying the quasilinear response relations to give a physical interpretation for the transitions in power and thermal efficiency. We also confirmed that the conceptual model can be obtained by averaging fast variables of the original model, and that unlike the quasilinear response relations, it preserves the thermodynamic irreversibility of the coupled Stirling engines. Furthermore, we compared the dynamics between the original and conceptual models for two-engine systems. Generalization of the theory developed in this study to a form applicable to a broader class of non-equilibrium autonomous heat engines is a challenging subject for future work.

\section*{Acknowledgements}

This work was supported by JSPS KAKENHI, Grant Numbers 21K12056 and 22K03450, and JST SPRING, Grant Number JPMJSP2108.

\section*{Appendix A: Derivation of Eq. (\ref{quasilinear})}

We derive the quasilinear response relations between thermodynamic fluxes and forces given by Eq. (\ref{quasilinear}). To this end, we take the time average on both sides of Eq. (\ref{dynamics}):
\begin{equation}\label{dynamics_averaging}
0=\sigma\Biggl\langle\left(\frac{T(\theta_i,\omega_i)}{V(\theta_i)}-P_{\rm air}\right)\sin\theta_i\Biggr\rangle-\Gamma\langle\omega_i\rangle-T_{\rm load}^{(i)}-\sum_{j\in\mathcal{N}_{i}}K_{ij}\Bigl\langle\sin\left(\theta_{i}-\theta_j\right)\Bigr\rangle.
\end{equation}
By expanding $T(\theta_i,\omega_i)$ w.r.t. $\omega_i$ as
\begin{equation}\label{temp_expansion}
T(\theta_i,\omega_i)=T_{{\rm eff}}(\theta_i)-\frac{\sigma\sin\theta_i}{GV}\omega_i+\mathcal{O}\left(\Delta T\omega_i, \omega_i^2\right).
\end{equation}
and substituting Eq. (\ref{temp_expansion}) into Eq. (\ref{dynamics_averaging}), we obtain
\begin{align}\label{dynamics_averaging_rewritten}
0=\sigma\Biggl\langle\left(\frac{T_{{\rm eff}}(\theta_i)}{V(\theta_i)}-P_{\rm air}\right)\sin\theta_i\Biggr\rangle-\frac{\sigma^2}{G}\Biggl\langle\frac{\sin^2\theta_i}{V^2(\theta_i)}\omega_i\Biggr\rangle
-\Gamma\langle\omega_i\rangle-T_{{\rm load}}^{(i)}-\sum_{j\in\mathcal{N}_{i}}K_{ij}\Bigl\langle\sin\left(\theta_{i}-\theta_j\right)\Bigr\rangle+\mathrm{H.O.T.}.
\end{align}
Let $\tau_k^{(i)}$ be the time required for $\theta_i$ to increase from $\theta_i(0)+2(k-1)\pi$ to $\theta_i(0)+2k\pi$ and define $\overline{\omega_k^{(i)}}$ as $\overline{\omega_k^{(i)}}\equiv2\pi/\tau_k^{(i)}$. The first and second terms on the left-hand side in Eq. (\ref{dynamics_averaging_rewritten}) can then be calculated as follows:
\begin{align}
&\sigma\Biggl\langle\left(\frac{T_{{\rm eff}}(\theta_i)}{V(\theta_i)}-P_{\rm air}\right)\sin\theta_i\Biggr\rangle\notag\\
&=\sigma\left(\lim_{N\to\infty}\frac{1}{\sum_{k=1}^N\tau_k^{(i)}}\sum_{l=1}^N\left(1\slash\overline{\omega_l^{(i)}}\right)\int_0^{\tau_l^{(i)}}\left(\frac{T_{{\rm eff}}(\theta_i)}{V(\theta_i)}-P_{\rm air}\right)\sin\theta_i\overline{\omega_l^{(i)}}dt\right)\label{omega_term1_1}\\
&=\frac{\sigma}{2\pi}\int_0^{2\pi}\left(\frac{T_{{\rm eff}}(\theta)}{V(\theta)}-P_{\rm air}\right)\sin\theta d\theta+\mathrm{H.O.T.}\\
&=\frac{\sigma}{2}\Biggl\langle\frac{\sin^2\theta}{V(\theta)}\Biggr\rangle_\theta\Delta T+\mathrm{H.O.T.}\label{dynamics_averaging_rewritten_first},
\end{align}
\begin{align}
\frac{\sigma^2}{G}\Biggl\langle\frac{\sin^2\theta_i}{V^2(\theta_i)}\omega_i\Biggr\rangle
&=
\frac{\sigma^2}{G}\left(\lim_{N\to\infty}\frac{1}{\sum_{k=1}^N\tau_k^{(i)}}\int_0^{2\pi N}\frac{\sin^2\theta}{V^2(\theta)}d\theta\right)\\
&=
\frac{\sigma^2}{G}\left(\frac{1}{2\pi}\int_0^{2\pi }\frac{\sin^2\theta}{V^2(\theta)}d\theta\right)\left(\lim_{N\to\infty}\frac{2\pi N}{\sum_{k=1}^N\tau_k^{(i)}}\right)\\
&=
\frac{\sigma^2}{G}\Biggl\langle\frac{\sin^2\theta}{V^2(\theta)}\Biggr\rangle_\theta\langle\omega_i\rangle\label{dynamics_averaging_rewritten_second}.
\end{align}
Substituting Eq. (\ref{dynamics_averaging_rewritten_first}) and Eq. (\ref{dynamics_averaging_rewritten_second}) into Eq. (\ref{dynamics_averaging_rewritten}), we obtain
\begin{equation}\label{omega_average}
\left(\Gamma+\frac{\sigma^2}{G}\Biggl\langle\frac{\sin^2\theta}{V^2(\theta)}\Biggr\rangle_\theta\right)\langle\omega_i\rangle=\frac{\sigma}{2}\Biggl\langle\frac{\sin^2\theta}{V(\theta)}\Biggr\rangle_\theta\Delta T-\-T_{{\rm load}}^{(i)}-\sum_{j\in\mathcal{N}_{i}}K_{ij}\Bigl\langle\sin\left(\theta_{i}-\theta_j\right)\Bigr\rangle+\mathrm{H.O.T.}.
\end{equation}
Neglecting higher-order terms, it is straightforward to obtain the quasilinear response relations w.r.t. $\langle\omega_{\rm m}\rangle$ and $\Bigl\langle\omega_{\rm d}^{(i)}\Bigr\rangle$. On the other hand, $\Bigl\langle J_{Q_{\rm b}}^{(i)}\Bigr\rangle$ can be written as
\begin{align}
\Bigl\langle J_{Q_{\rm b}}^{(i)}\Bigr\rangle&=\Biggl\langle G\frac{1+\sin\theta_i}{2}(T_{\rm b}-T(\theta_i,\omega_i))\Biggr\rangle\\
&=
\Biggl\langle G\frac{1+\sin\theta_i}{2}(T_{\rm b}-T_{{\rm eff}}(\theta_i))\Biggr\rangle
+\Biggl\langle\frac{1+\sin\theta_i}{2}\frac{\sigma\sin\theta_i}{V(\theta)}\omega_i\Biggr\rangle+\mathrm{H.O.T.}\label{heat flux}.
\end{align}
The first and second terms in Eq. (\ref{heat flux}) can then be calculated in the same way as in Eqs. (\ref{omega_term1_1})-(\ref{dynamics_averaging_rewritten_second}) and are obtained as $\displaystyle\frac{G}{8}\Delta T+\mathrm{H.O.T.}$ and $\displaystyle\frac{\sigma}{2}\Biggl\langle\frac{\sin^2\theta}{V(\theta)}\Biggr\rangle_\theta\langle\omega_i\rangle$, respectively. Neglecting higher-order terms, the quasilinear response relation w.r.t. $\displaystyle\sum_{i=1}^N\Bigl\langle J_{Q_{\rm b}}^{(i)}\Bigr\rangle$ can be obtained straightforwardly.

\section*{Appendix B: Analysis based on the energy conservation law}
In this section, we calculate the averaged brake power $\langle P_{\mathrm{load}}\rangle$ of the coupled Stirling engines using the averaged heat flux equation and the energy conservation law to confirm that it agrees with the result obtained from the equation of motion of the conceptual model constructed in Section III.

From Eq. (\ref{J_b_ave_first}), the averaged equation for the heat flux from the heat reservoir of temperature $T_{\mathrm b}$ is given by
\begin{align}
\sum_{i=1}^N J_{Q_{\mathrm b}}^{(i)}(\theta_i,\omega_i)\approx\frac{G}{8}N\Delta T+\frac{N\sigma}{2}\Biggl\langle\frac{\sin^2\theta}{V(\theta)}\Biggr\rangle_\theta\omega_i.
\end{align}
On the other hand, the averaged equation for the heat flux from the heat reservoir of temperature $T_{\mathrm t}$ is also given by Eq. (\ref{J_b_ave_first}), indicating that the sum of the heat fluxes from the two heat reservoirs is zero. Therefore, analysis using the energy conservation law requires the approximation of the heat flux to the second order for minute quantities. To this end, we preserve the temperature $T(\theta_i,\omega_i)$ to the second order for minute quantities instead of truncating $\mathcal{O}\left(\Delta T\omega_i, \omega_i^2\right)$ as in Eq. (\ref{temp_expansion}):
\begin{equation}\label{temp_expansion_secondorder}
T(\theta_i,\omega_i)\approx T_{{\mathrm{eff}}}(\theta_i)-\frac{\sigma\sin\theta_i}{GV}\omega_i-\frac{\sigma\sin^2\theta_i}{2GV(\theta_i)}\Delta T\omega_i+\left(\frac{\sigma\sin\theta_i}{GV(\theta_i)}\right)^2\omega_i^2.
\end{equation}
Substituting Eq. (\ref{temp_expansion_secondorder}) into Eq. (\ref{heatflux}) and averaging fast variables as in subsection III. C, the heat fluxes $J_{Q_{\mathrm b}}^{(i)}(\theta_i,\omega_i)$ and $J_{ Q_{\mathrm t}}^{(i)}(\theta_i,\omega_i)$ can be obtained as follows:
\begin{align}
J_{Q_{\mathrm b}}^{(i)}(\theta_i,\omega_i)\approx\frac{G}{8}\Delta T+\frac{\sigma}{2}\Biggl\langle\frac{\sin^2\theta}{V(\theta)}\Biggr\rangle_\theta\omega_i+\frac{\sigma}{4}\Biggl\langle\frac{\sin^2\theta}{V(\theta)}\Biggr\rangle_\theta\Delta T\omega_i-\frac{\sigma^2}{2G}\Biggl\langle\frac{\sin^2\theta}{V^2(\theta)}\Biggr\rangle_\theta\omega_i^2,
\end{align}
\begin{align}
J_{Q_{\mathrm t}}^{(i)}(\theta_i,\omega_i)\approx-\frac{G}{8}\Delta T-\frac{\sigma}{2}\Biggl\langle\frac{\sin^2\theta}{V(\theta)}\Biggr\rangle_\theta\omega_i+\frac{\sigma}{4}\Biggl\langle\frac{\sin^2\theta}{V(\theta)}\Biggr\rangle_\theta\Delta T\omega_i-\frac{\sigma^2}{2G}\Biggl\langle\frac{\sin^2\theta}{V^2(\theta)}\Biggr\rangle_\theta\omega_i^2.
\end{align}
Thus, the averaged equation for the sum of the heat fluxes from the two heat reservoirs is given by
\begin{align}
&\sum_{i=1}^N\left[J_{Q_{\mathrm b}}^{(i)}(\theta_i,\omega_i)+J_{Q_{\mathrm t}}^{(i)}(\theta_i,\omega_i)\right]\\
&\approx\sum_{i=1}^N\left[\frac{\sigma}{2}\Biggl\langle\frac{\sin^2\theta}{V(\theta)}\Biggr\rangle_\theta\Delta T\omega_i-\frac{\sigma^2}{G}\Biggl\langle\frac{\sin^2\theta}{V^2(\theta)}\Biggr\rangle_\theta\omega_i^2\right]\\
&=\frac{N\sigma}{2}\Biggl\langle\frac{\sin^2\theta}{V(\theta)}\Biggr\rangle_\theta\Delta T\omega_{\mathrm m}-\frac{N\sigma^2}{G}\Biggl\langle\frac{\sin^2\theta}{V^2(\theta)}\Biggr\rangle_\theta\omega_{\mathrm m}^2-\frac{\sigma^2}{G}\Biggl\langle\frac{\sin^2\theta}{V^2(\theta)}\Biggr\rangle_\theta\left[\sum_{i=1}^N\left(\omega_{\mathrm d}^{(i)}\right)^2\right].\label{J_bt_secondorder}
\end{align}
On the other hand, since the power dissipated by viscous friction $\displaystyle\sum_{i=1}^N P_{\mathrm{fric}}^{(i)}$ is given by
\begin{align}
\sum_{i=1}^N P_{\mathrm{fric}}^{(i)}=\sum_{i=1}^N\Gamma\omega_i^2=\Gamma\left[N\omega_{\mathrm m}^2+\sum_{i=1}^N\left(\omega_{\mathrm d}^{(i)}\right)^2\right],
\end{align}
the averaged brake power $\langle P_{\mathrm{load}}\rangle$ can be obtained from the energy conservation law and Eq. (\ref{J_bt_secondorder}):
\begin{align}
\langle P_{\mathrm{load}}\rangle
&=\sum_{i=1}^N\biggl\langle J_{Q_{\mathrm b}}^{(i)}(\theta_i,\omega_i)+J_{Q_{\mathrm t}}^{(i)}(\theta_i,\omega_i)\biggr\rangle-\sum_{i=1}^N \Bigl\langle P_{\mathrm{fric}}^{(i)}\Bigr\rangle\\
&\approx \left[\frac{N\sigma}{2}\Biggl\langle\frac{\sin^2\theta}{V(\theta)}\Biggr\rangle_\theta\Delta T\langle\omega_{\mathrm m}\rangle-N\left(\Gamma+\frac{\sigma^2}{G}\Biggl\langle\frac{\sin^2\theta}{V^2(\theta)}\Biggr\rangle_\theta\right)\bigl\langle\omega_{\mathrm m}^2\bigr\rangle\right]-\left(\Gamma+\frac{\sigma^2}{G}\Biggl\langle\frac{\sin^2\theta}{V^2(\theta)}\Biggr\rangle_\theta\right)\left(\sum_{i=1}^N\biggl\langle\left(\omega_{\mathrm d}^{(i)}\right)^2\biggr\rangle\right).\label{P_ave_higher_J}
\end{align}
The first term in Eq. (\ref{P_ave_higher_J}) coincides with the averaged brake power $\displaystyle \Bigl\langle P_{\mathrm{load}}^{\mathrm M}\Bigr\rangle$ of the engine ${\mathrm M}_{\mathrm{EG}}$ in system M, which can be confirmed by multiplying both sides of Eq. (\ref{dynamics_M}) by $N\omega_{\mathrm m}$ and averaging it over time. On the other hand, the second term in Eq. (\ref{P_ave_higher_J}) agrees with the averaged brake power $\Bigl\langle P_{\mathrm{load}}^{\mathrm D}\Bigr\rangle$ for system D. Therefore, Eq. (\ref{P_ave_higher_J}) agrees with the averaged brake power corresponding to the conceptual model constructed in III. B. It can be seen that the power dissipated by the viscous friction of the conceptual model consists of the power dissipated by the viscous friction due to the rotational motion of the Stirling engines, $\displaystyle\sum_{i=1}^N\Gamma\Bigl\langle\omega_i^2\Bigr\rangle$, and the loss of the energy flow that the Stirling engines receive from the heat reservoirs, $\displaystyle\frac{\sigma^2}{G}\biggl\langle\frac{\sin^2\theta}{V^2(\theta)}\biggr\rangle_\theta\left(\sum_{i=1}^N\Bigl\langle\omega_i^2\Bigr\rangle\right)$. The latter is the heat leakage due to nonquasi-static processes associated with the rotational motion of the Stirling engines, which depends on the magnitude of the coupling strength: the portion of the heat leakage that depends on the motion of the mean angle of the coupled Stirling engines, $\displaystyle\frac{\sigma^2}{G}\biggl\langle\frac{\sin^2\theta}{V^2(\theta)}\biggr\rangle_\theta\left(\sum_{i=1}^N\Bigl\langle\omega_{\mathrm m}^2\Bigr\rangle\right)$, is kept at a constant value, while the remaining that depends on the relative motion, $\displaystyle\frac{\sigma^2}{G}\biggl\langle\frac{\sin^2\theta}{V^2(\theta)}\biggr\rangle_\theta\left(\sum_{i=1}^N\biggl\langle\left(\omega_{\mathrm d}^{(i)}\right)^2\biggr\rangle\right)$, becomes smaller when the coupling strength is increased. Thus, the conceptual model constructed in III. B interprets the heat leakage due to nonquasi-static processes as part of energy dissipation due to viscous friction in the conceptual model, thereby allowing the original coupled system to be seperated into two isolated subsystems: a coupling strength-independent system M and a weakly-coupled system D.

\bibliographystyle{unsrt}

\newpage

\end{document}